\documentclass[letterpaper]{article}

\usepackage{graphicx}
\graphicspath{{./Fig/}}
\usepackage{color}
\usepackage{placeins}
\usepackage{float}
\usepackage{tabularx,colortbl}
\usepackage{amsmath}
\usepackage{amsfonts}
\usepackage{caption}
\usepackage{soul}
\usepackage{authblk}
\usepackage{amssymb}
\usepackage{amsmath}
\usepackage{amsthm}
\usepackage{amsfonts}
\usepackage{authblk}
\usepackage{indentfirst}
\usepackage{subfig}
\usepackage{mathtools}
\usepackage{bm}
\usepackage{mathrsfs}
\usepackage{fancyhdr}
\usepackage{enumerate}
\usepackage{booktabs}
\usepackage{colortbl}
\usepackage{multicol}
\usepackage{multirow}
\usepackage{cancel,soul}
\setstcolor{red}
\usepackage{xcolor}
\usepackage{nth}
\usepackage{url}

\usepackage{stfloats}
\usepackage[font=small,skip=0pt]{caption}
\usepackage[boxed,linesnumbered]{algorithm2e}
\usepackage{listings}
\providecommand{\keywords}[1]{\textbf{\textit{Keywords---}} #1}
\newcommand{\argmin}{\operatornamewithlimits{argmin}}

\newcommand{\real}{\mathbb{R}}

\newcommand{\bn}{\mathbf{n}}
\newcommand{\bp}{\mathbf{p}}
\newcommand{\bx}{\mathbf{x}}

\newcommand{\bw}{\mathbf{w}}

\newcommand{\bA}{\mathbf{A}}

\newcommand{\ba}{\mathbf{a}}

\newcommand{\bd}{\mathbf{d}}

\newcommand{\bfm}{\mathbf{m}}

\newcommand{\bD}{\mathbf{D}}

\newcommand{\bbs}{\mathbf{s}}

\newcommand{\bX}{\mathbf{X}}
\newcommand{\bY}{\mathbf{Y}}

\newcommand{\bPhi}{\mathbf{\Phi}}

\title{Joint Seismic Data Denoising and Interpolation with Double-Sparsity Dictionary Learning}

\author[1]{Lingchen Zhu\thanks{lczhu@gatech.edu}}
\author[1]{Entao Liu\thanks{liuentao@gmail.com}}
\author[1]{James H. McClellan\thanks{jim.mcclellan@ece.gatech.edu}}
\affil[1]{CeGP at Georgia Institute of Technology}

\begin{document}
\maketitle

\renewcommand{\thefootnote}{\fnsymbol{footnote}}

%% abstract
 \begin{abstract}
Seismic data quality is vital to geophysical applications, so methods of data recovery, including denoising and interpolation, are common initial steps in the seismic data processing flow.
We present a method to perform simultaneous interpolation and denoising, which is based on double-sparsity dictionary learning.
This extends previous work that was for denoising only.
The original double sparsity dictionary learning algorithm is modified to track the traces with missing data by defining a masking operator that is integrated into the sparse representation of the dictionary.
A weighted low-rank approximation algorithm is adopted to handle the dictionary updating as a sparse recovery optimization  problem constrained by the masking operator. 
Compared to traditional sparse transforms with fixed dictionaries that lack the ability to adapt to complex data structures, the double-sparsity dictionary learning method learns the signal adaptively from selected patches of the corrupted seismic data while preserving compact forward and inverse transform operators.
Numerical experiments on synthetic seismic data indicate that this new method preserves more subtle features in the dataset without introducing pseudo-Gibbs artifacts when compared to other directional multiscale transform methods such as curvelets.
\end{abstract}

\begin{figure}[h!]
	\centering
	\subfloat[Original seismic dataset]
	{
		\begin{minipage}{0.4\linewidth}
			\centering
			\includegraphics[width=\textwidth]{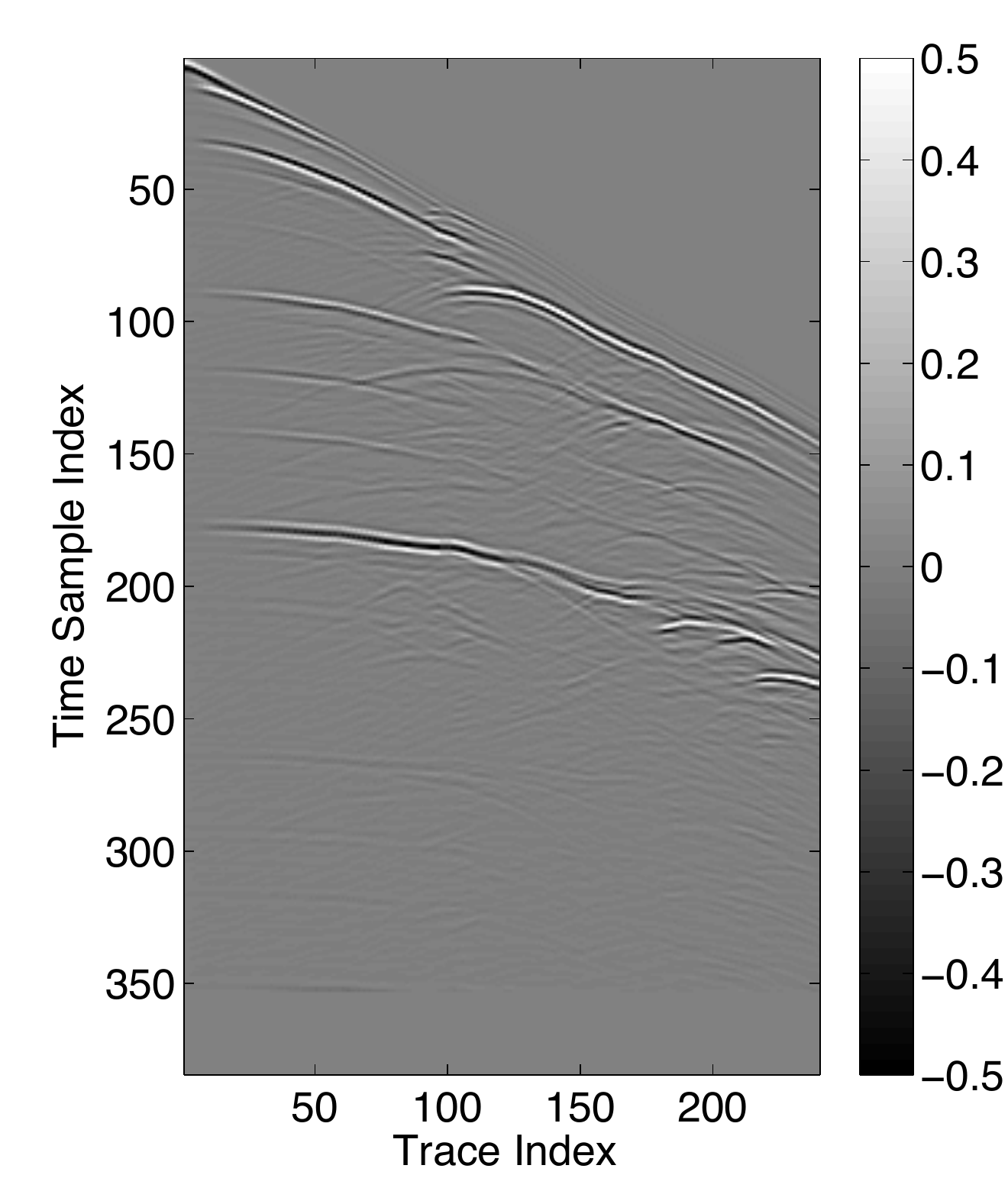}
		\end{minipage}
	}\\
	\subfloat[Noisy dataset, 33\% missing traces]
	{
		\begin{minipage}{0.4\linewidth}
			\centering
			\includegraphics[width=\textwidth]{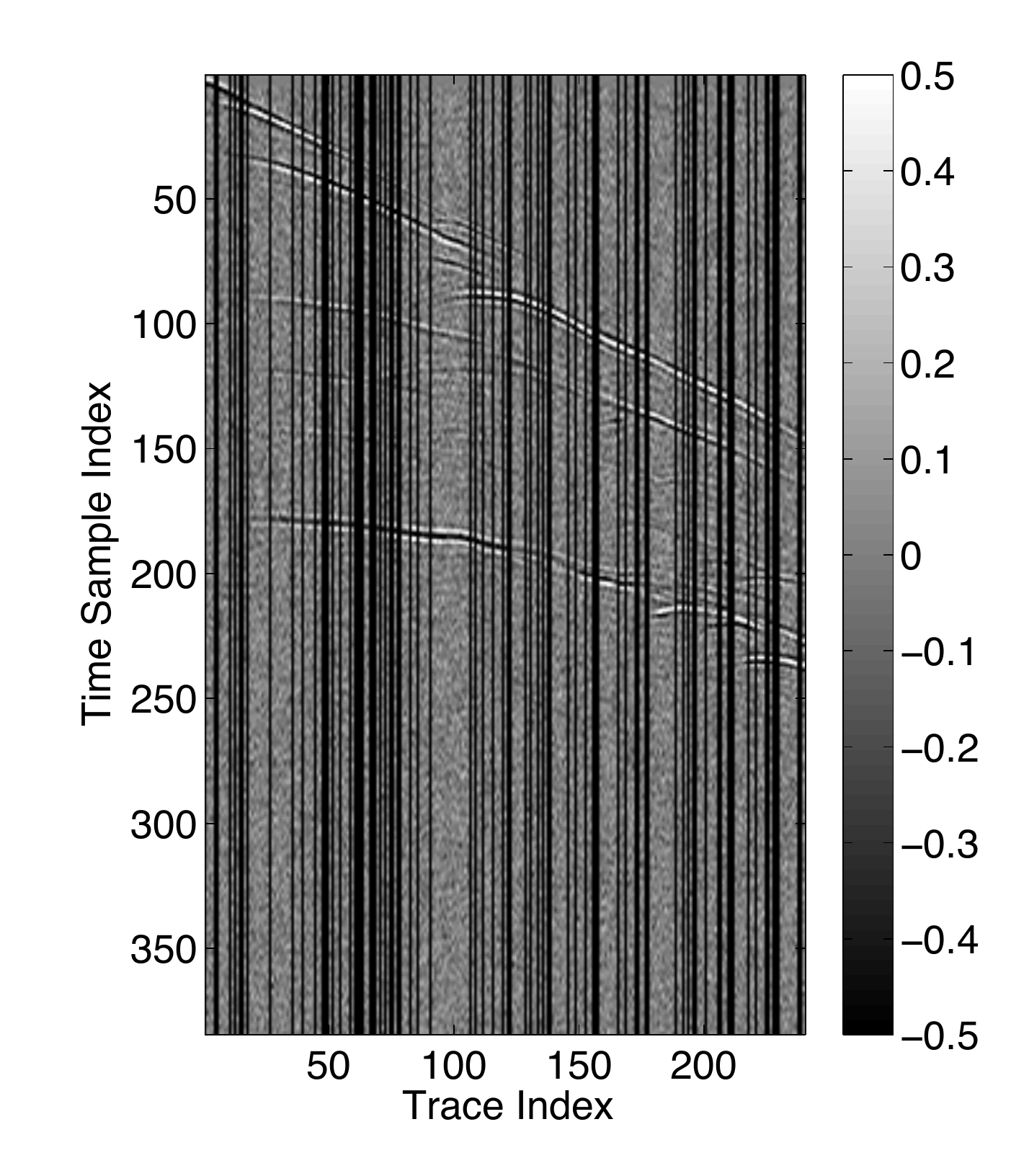}
		\end{minipage}
		\label{fig:bppublic_noisyData_forInpainting_033}
	}
	\subfloat[Noisy dataset, 50\% missing traces]
	{
		\begin{minipage}{0.4\linewidth}
			\centering
			\includegraphics[width=\textwidth]{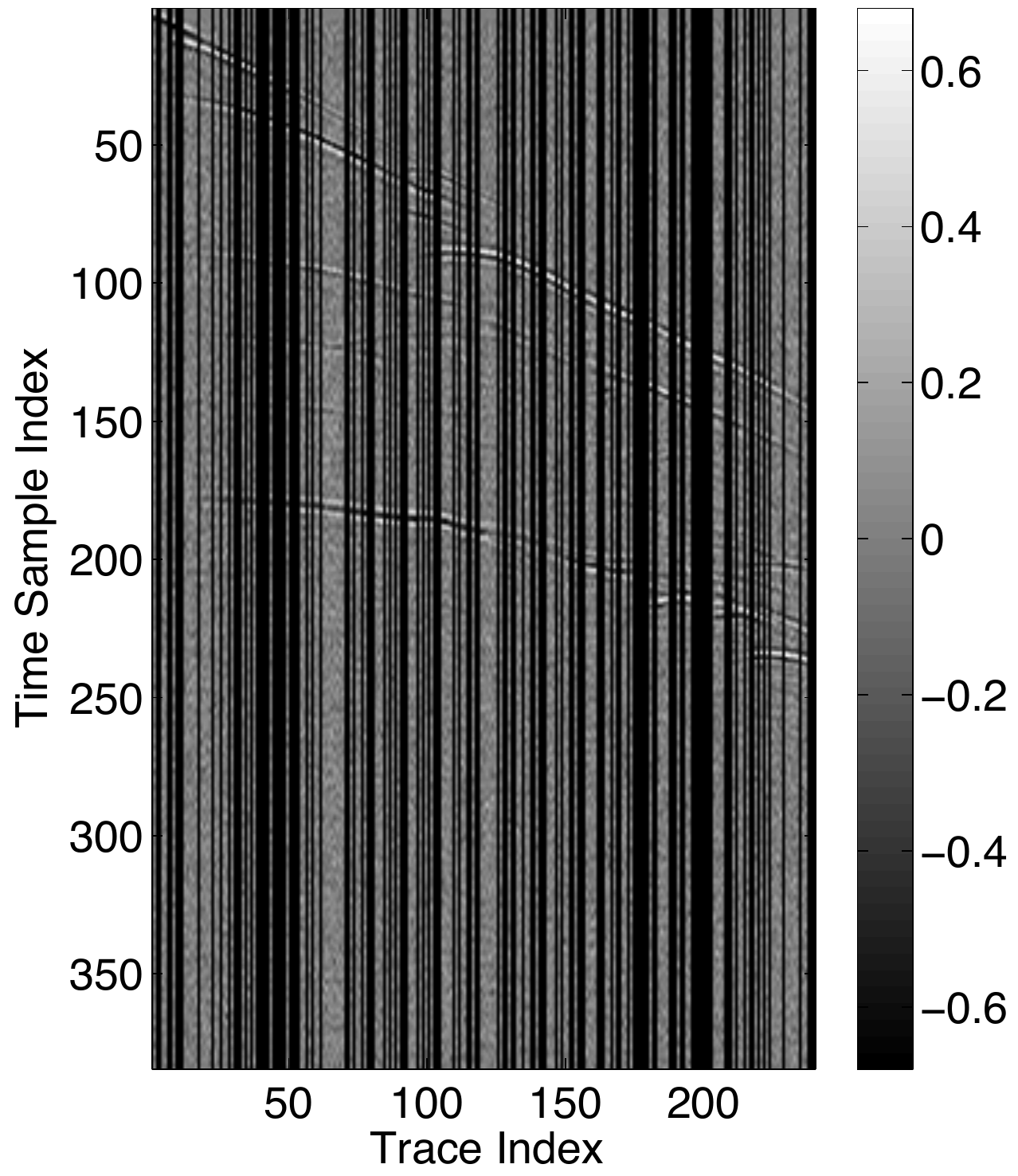}
		\end{minipage}
		\label{fig:bppublic_noisyData_forInpainting_050}
	}
	\caption{BP dataset, original, and noisy $(\sigma = 0.1)$  with missing traces (shown as black traces).}
	\label{fig:bppublic_data_for_inpainting}
\end{figure}

A seismic dataset is an ensemble of time-domain wiggle traces collected from an array of receivers. In exploration geophysics a seismic wavefield recording is processed to produce estimates of various properties of the Earth's subsurface. However, the recorded data may suffer not only from correlated and uncorrelated noise, but also from missing traces due to various constraints (see Figure \ref{fig:bppublic_data_for_inpainting} for examples), such as malfunctioning sensors, limited budget, lack of permission to access the field, etc. Many seismic applications, such as event detection, migration, and inversion have strict requirements on the quality of data. Preconditioning of the recorded data is typically needed for those applications to yield satisfactory results. In this work we concentrate on two major tasks: denoising that attenuates the noise and interpolation that reconstructs the missing traces, and propose a scheme based on dictionary learning that fulfills both of these goals. For 2D seismic data, the interpolation task is actually equivalent to a special case of inpainting in imaging processing.

% literature survey
In recent years, sparse representation of seismic signals via transform-domain methods has attracted considerable attention in seismic data recovery. This model suggests that natural seismic signal are compressible, or well approximated, by a linear combination of only a few atoms from a dictionary. Imposing sparsity constraints on the coefficients of the linear combination could efficiently eliminate the anomalies in the signal recovered from the contaminated seismic data. Good results in both denoising and interpolation have been reported using different transforms/dictionaries, such as wavelets \cite{Zhang:2003aa}, curvelets \cite{Hennenfent:2006aa, Herrmann:2008aa}, contourlets \cite{Do:2005aa}, seislets \cite{Fomel:2010aa}, etc.

%% regular DL needs a training database which usually is not available
The transform-domain methods above assume specific underlying regularity of the data described by analytic models, resulting in transforms according to implicit but fixed dictionaries that process the data section as a whole. Alternatively, data driven sparse dictionaries \cite{Zhu:2015aa} or tight frames \cite{YuMaAO2015} can be learned directly from the dataset. This ad hoc learned dictionary, which is typically in the form of explicit matrices for small patches instead of full scale dataset, can better adapt the complex data characteristics. For instance, better denoising results obtained using double sparse dictionary learning rather than generic transforms such as curvelet, contourlet, or seislet have been reported in \cite{Zhu:2015aa,ChenEtAl2016}. As an effective dictionary learning algorithm, K-Singular Value Decomposition (K-SVD) \cite{Aharon:2006aa} has been adopted in seismic denoising problems \cite{Tang:2012aa, Beckouche:2014aa}. However, the major drawback of K-SVD is its high computational complexity. To this end, in this paper we propose to apply the efficient double sparsity dictionary learning approach \cite{Rubinstein:2010aa}, which further squeezes out the redundancy in atoms of the learned dictionary.

This paper is organized as follows. In Section \ref{sec:DSDL} we revisit the double sparsity dictionary learning. Then we provide the scheme to solve the joint denoising and interpolation problem in Section \ref{sec:DLMD}. In Section \ref{sec:numerical} we give numerical simulations of the proposed scheme and Section \ref{sec:conclusions} concludes this paper.

\section{Double Sparsity Dictionary Learning}\label{sec:DSDL}
%Let us recap the double sparsity dictionary learning before introducing the mask operator.
Given a training set $\mathbf{Y} = [\mathbf{y}_1, \mathbf{y}_2, \dots, \mathbf{y}_R] \in \mathbb{R}^{N \times R}$ that contains $R$ training samples, the dictionary learning process looks for $\mathbf{D}=[\bd_1,\ldots,\bd_L] \in \mathbb{R}^{N \times L}$ such that $\bY \approx \bD \bX$, where $\mathbf{X} = [\mathbf{x}_1, \mathbf{x}_2, \dots, \mathbf{x}_R] \in \mathbb{R}^{L \times R}$ is a coefficient matrix that is sparse. 
Here we assume atoms in $\bD$ are normalized without loss of generality, i.e., $\forall j\;\|\bd_j\|_2=1$, and $N \leq L$.
This dictionary learning process could be accomplished by solving the following tractable convex $\ell_1$ minimization problem:
\begin{eqnarray}
\label{eq:ksvd-learning-l1}
\{\hat{\mathbf{D}}, & \hat{\mathbf{X}}\} = \argmin\limits_{\mathbf{D}, \mathbf{X}} \|\mathbf{Y} - \mathbf{DX}\|_F^2 \quad \mbox{s.t. } \|\mathbf{x}_i\|_1 \leq t, \quad \forall i = 1, \dots, R,
\end{eqnarray}
where $\|\cdot\|_F$ is the Frobenius norm. This $\ell_1$-norm minimization problem is relaxed from the $\ell_0$-norm minimization, which is well known to be NP-hard and can not be solved directly. 
Under certain conditions \cite{Donoho:2006aa}, problem (\ref{eq:ksvd-learning-l1}) yields an exact solution to the $\ell_0$-norm problem. Hereafter, we use the $\ell_1$-norm to measure the sparsity level.
%For large scale problems, such as natural images and seismic data sections, it is prohibitively expensive to train the learned dictionary using all of the data. Moreover, the large number of training samples required by dictionary learning algorithms is usually not available in the denoising and interpolation applications. Therefore, we follow the convention of K-SVD and divide the seismic data into many small patches to mitigate the computation burden. Since non-overlapping patches usually result in artifacts on their boundaries, we use overlapping patches which have better consistency in capturing the underlying data features as well as less blocking artifacts.
Here we follow the convention of K-SVD and divide the seismic data into many small overlapping patches to mitigate the computation burden, the advantages of this approach over the training using whole data is discussed in \cite{Zhu:2015aa}.

Because of the physical properties of seismic wave propagation and reflectivity, atoms with similar geometric features are commonly observed in learned dictionaries \cite{Zhu:2015aa}. Therefore, it is efficient to employ an off-the-shelf base dictionary to  represent the atoms in the learned dictionary. The form we choose is
\begin{equation}
\label{eq:dict-double-sparsity}
\mathbf{D} = \mathbf{\Phi A},
\end{equation}
where $\mathbf{\Phi} \in \mathbb{R}^{N \times L}$ is a fixed base dictionary and $\mathbf{A} = [\mathbf{a}_1, \mathbf{a}_2, \dots, \mathbf{a}_L] \in \mathbb{R}^{L \times L}$ is a sparse matrix to be learned in which each column satisfies $\|\mathbf{a}_i\|_1 \leq p$ for some sparsity level $p$. 
%{\CRED (Intuitively? or Empirically?)}, 
Intuitively, a base dictionary $\bPhi$ that fits well to the expected data regularities should give better final outputs.
The advantage of the model in (\ref{eq:dict-double-sparsity}) is that the output of dictionary learning is reduced from finding all the entries in a full matrix $\bD$ to finding the sparse matrix $\mathbf{A}$, which significantly impacts the efficiency of computation, storage, and transmission. More importantly, with fewer degrees of freedom, such a dictionary model reduces the chance of overfitting the noise in the training set and produces robust results even with a limited number of training examples. For a fixed generic base dictionary $\bPhi$, the dictionary learning process for $\bD$ in (\ref{eq:dict-double-sparsity}) becomes
\begin{equation}
\label{eq:sparse-ksvd-learning}
\{\hat{\mathbf{A}}, \hat{\mathbf{X}}\} = \argmin\limits_{\mathbf{A}, \mathbf{X}} \|\mathbf{Y} - \mathbf{\Phi AX}\|_F^2\nonumber
\end{equation}
\begin{equation}
\mbox{ s.t. }\begin{cases}
\|\mathbf{x}_i\|_1 \leq t, & \forall i = 1, \dots, R\\
\|\mathbf{a}_j\|_1 \leq p, \|\mathbf{\Phi a}_j\|_2 = 1 & \forall j = 1, \dots, L.
\end{cases}
\end{equation}
Because both the actual learned dictionary $\mathbf{A}$ and the representation coefficients $\mathbf{X}$ are sparse matrices, this model is called \emph{double sparsity} dictionary learning \cite{Rubinstein:2010aa} and (\ref{eq:sparse-ksvd-learning}) can be solved by the sparse K-SVD  algorithm.
To be honest, the name sparse K-SVD is a bit misleading, since the columns in $\bA$ are updated one at a time using sparse coding rather than by rank-1 approximation as in the original K-SVD method \cite{Aharon:2006aa}.
To deal with the optimization involving two arguments, the authors of \cite{Rubinstein:2010aa} developed a scheme that alternates the updating of $\bA$ or $\bX$ while keeping the other one fixed.

We now adapt their approach to the missing traces scenario.
When updating $\ba_j$, one column of $\mathbf{A}$, we determine the column index set $\mathcal{I}_j$ of the training samples in $\mathbf{Y}$ whose representations use $\mathbf{d}_j=\bPhi\ba_j$ which are found from nonzero entries in $\mathbf{X}$
\begin{equation}
\mathcal{I}_j = \{r|\; 1\le r \le R, x_{jr}\not=0 \}.
\end{equation}
Then the objective functional for $\mathbf{a}_j$ in (\ref{eq:sparse-ksvd-learning}) can be written as
\begin{eqnarray}
\label{eq:sparse-ksvd-atom-update}
\left\| \mathbf{Y}_{\mathcal{I}_j} - \mathbf{\Phi AX}_{\mathcal{I}_j} \right\|_F^2 
&= \biggl\|\biggl(\mathbf{Y}_{\mathcal{I}_j} - \sum\limits_{i \neq j}\mathbf{\Phi a}_i\mathbf{X}_{i,\mathcal{I}_j}\biggr) - \mathbf{\Phi a}_j\mathbf{X}_{j,\mathcal{I}_j}\biggr\|_F^2\nonumber\\
&= \left\| \mathbf{E}_j - \mathbf{\Phi a}_j\mathbf{X}_{j,\mathcal{I}_j} \right\|_F^2,
\end{eqnarray}
where $\mathbf{E}_j = \mathbf{Y}_{\mathcal{I}_j} - \sum\limits_{i \neq j}\mathbf{\Phi a}_i\mathbf{X}_{i,\mathcal{I}_j}$ is the residual matrix without the contribution of $\mathbf{d}_j$. Therefore, the resulting problem to update $\mathbf{a}_j$ and $\mathbf{X}_{j,\mathcal{I}_j}$ is given by
\begin{equation}
\label{eq:sparse-ksvd-rank1-approximation}
\left\{\mathbf{a}_j, \mathbf{X}_{j,\mathcal{I}_j}^T\right\} = \argmin\limits_{\mathbf{a}, \mathbf{x}} \left\| \mathbf{E}_j - \mathbf{\Phi ax}^T \right\|_F^2 \quad \mbox{s.t. }
\begin{cases}
	\|\mathbf{a}\|_1  \leq p\\
	\|\mathbf{\Phi a}\|_2 = 1.\\
	\end{cases}
\end{equation}
By virtue of Lemma 1 in \cite{Rubinstein:2010aa}, (\ref{eq:sparse-ksvd-rank1-approximation}) is equivalent to the following sparse coding problem:
\begin{eqnarray}
\label{eq:simpleform}
\left\{\mathbf{a}_j, \mathbf{X}_{j,\mathcal{I}_j}^T\right\} = &\argmin\limits_{\mathbf{a}, \mathbf{x}} \left\| \mathbf{E}_j \bx - \mathbf{\Phi a} \right\|_F^2 \quad \mbox{s.t. }
\begin{cases}
	\|\mathbf{a}\|_1 \leq p\\
	\|\mathbf{\Phi a}\|_2 = 1.\\
	\end{cases}
\end{eqnarray}
which updates  only those columns within the index set $\mathcal{I}_j$ at the $j^{\mbox{th}}$ row of $\mathbf{X}$.
%{\CRED Lingchen:  Aren't we updating the entire $j^{\mbox{th}}$ row of $\mathbf{X}$ when we iterate in \eqref{eq:simpleform} ?  i.e., use $\mathbf{x}^T_j$ instead of $\mathbf{X}_{j,\mathcal{I}_j}^T$}\\
%{\mboxcolor{blue}{Lingchen's Comment: Not the entire $j^{\mbox{th}}$ row of $\mathbf{X}$, but only those columns within index set $\mathcal{I}_j$ at the $j^{\mbox{th}}$ row, as the atom $\mathbf{a}_j$ has already been canceled and only $\mathbf{X}_{j,\mathcal{I}_j}$ is affected. This is a row vector denoted as $\mathbf{X}_{j,\mathcal{I}_j}$, and its transposed vector $\mathbf{X}_{j,\mathcal{I}_j}^T$ is the solution of \eqref{eq:simpleform}. More importantly, the dictionary atom update step does not aim to update one entire row of $\mathbf{X}$ since doing so will sabotage the sparsity of $\mathbf{X}$. This step only updates non-zero elements (with index $\mathcal{I}_j$ meaning that they have used the atom $\mathbf{a}_j$ for representation) at the $j^{\mbox{th}}$ row of $\mathbf{X}$.}}

\section{Simultaneous Seismic Data Denoising and Interpolation}\label{sec:DLMD}
Besides random noise, the dictionary learning method is applicable to data distortion as well. For instance, distortions caused by missing traces and additive noise can be handled jointly using dictionary learning. The interpolation of  missing traces is crucial for many seismic applications, because inadequate or irregularly spaced traces in the acquired seismic dataset could produce strong artifacts in the following seismic processing stages. In practice, trace interpolation, along with denoising, has become one essential step in industrial seismic data preprocessing workflows.

Previously, a variety of methods have been developed for seismic dataset interpolation. Early work proposed a trace interpolation method by wave-equation methods based on the principles of wave physics \cite{Ronen:1987aa}. Later, methods based on the Fourier transform \cite{DuiSch1999,Liu:2004aa,Zwartjes:2007aa} were adopted to reconstruct irregularly sampled seismic signals. In the past decade, multi-scale transform methods have been widely used to fill gaps among traces based on the sparsity of seismic wave fronts in the transform domain \cite{Zhang:2003aa,Hennenfent:2006aa,Hennenfent:2008aa,Herrmann:2008ab,Naghizadeh:2010aa,Fomel:2010aa}. These methods process the dataset as a whole.

To start with we consider a small $n_z \times n_x$ patch in the 2D seismic data section which is then vectorized into $\bp \in \real^{N}$, where $N = n_z n_x$. The noisy patch $\bp$ is composed of the signal part $\bbs$ and noise $\bn$, i.e., $\bp =\bbs+\bn$.
%\begin{equation}
%\bp =\bbs+\bn.
%\end{equation}
For a given dictionary $\bD = \bPhi\bA$, denoising can be achieved by:
\begin{equation}
\{\hat\bbs,\hat\bx\} = \argmin_{\bbs,\bx}\|\bbs-\bD\bx\|_2^2 +\mu\|\bx\|_1 +\lambda\|\bbs-\bp\|_2^2,
\end{equation}
where $\mu$ and $\lambda$ control the balance among the above three terms: fidelity of the denoised result to the sparse model, sparsity level, and close fit to the original data.
For the general case, where $\bw\in\real^{N_z N_x}$ is the vector representation of the entire noisy seismic section, we define a patching operator as $\bm{\mathcal{R}}_{i}\bw = \bp_i\in\real^N$ which takes values from the $i^{\mbox{th}}$ noisy patch and then reshapes them into a vector.
In terms of the patches, the denoising problem for the whole seismic data section is written as:
%{\CRED Don't we need $\mu_i$ in this equation?
%	Should the LHS be $\hat\bx$ instead of $\hat\bx_i$ ?}\\
%{\mboxcolor{blue}{Lingchen's Comment: The subscript $i$ denotes the index of the patch. Different patches may have different noise upper bound due to the different distribution of missing traces, as shown in \eqref{eq:dataset-inpainting-sparse-representation}.}}
\begin{equation}\label{DN_whole}
\{\hat\bbs,\hat\bx_i\} = \argmin_{\bbs,\bx_i} \sum_{i} \|\bm{\mathcal{R}}_i \bbs - \bD\bx_i\|_2^2
+\sum_i \mu_i \|\bx_i\|_1 + \lambda \|\bbs-\bw\|^2_2.
\end{equation}
Again, a local optimal solution can be obtained by the alternating optimization over $\bbs$ and $\bx_i$. When $\bbs$ is fixed (\ref{DN_whole}) is a sparse coding problem, when $\bbs$ is fixed, equation (\ref{DN_whole}) is a sparse coding problem, and when $\bx_i$ is fixed it simplifies into a Tikhonov-regularized least squares problem which has a closed-form solution. 

In denoising problems, dictionary learning approaches typically employ all the available data to train the dictionary. However, when doing interpolation we have to keep track of missing data during the learning process to avoid artifacts, so the basic assumption is that the locations of all missing data are known. Thus, we use a mask operator in the learning process to mute those missing traces. The mask vector is denoted by $\bfm\in \{0,1\}^{N_z N_x}$ whose elements are
\begin{equation}
\label{eq:beta_value}
m_i = \begin{cases}
	1,& i^{\mbox{th}} \text{element corresponds to available data}\\
	0,&  i^{\mbox{th}} \text{element corresponds to a missing trace}.
	\end{cases}
\end{equation}
With $\odot$ denoting element-wise multiplication between two matrices or two vectors, the joint denoising and interpolation optimization problem becomes
\begin{eqnarray}
\label{eq:fused-dataset-learning-inpainting}
\left\{\hat{\mathbf{s}}, \hat{\mathbf{A}}, \hat{\mathbf{x}}_{i}\right\} =& \argmin_{\mathbf{s}, \mathbf{A}, \mathbf{x}_i} \sum_i \|(\bm{\mathcal{R}}_{i}\bfm) \odot (\bm{\mathcal{R}}_{i}\mathbf{s} - \bPhi \bA\bx_i)\|_2^2 \nonumber\\
& + \sum_{i} \mu_{i}\|\mathbf{x}_i\|_1+ \lambda \|\bfm \odot (\mathbf{s} - \mathbf{w})\|_2^2,
\end{eqnarray}
where the minimization over $\bA$ incorporates the dictionary learning step as well.

There are three alternating steps to solve this optimization problem. After initializing $\hat{\mathbf{s}} = \mathbf{w}$ and using a fixed $\mathbf{A}$, the sparse representation basis pursuit denoising (BPDN) \cite{BergFriedlander:2008} problem for each patch $\bm{\mathcal{R}}_{i}\hat{\mathbf{s}}$ becomes
\begin{eqnarray}
\label{eq:dataset-inpainting-sparse-representation}
&\mathbf{x}_{i} = \argmin\limits_{\mathbf{x}} \|\mathbf{x}\|_1\nonumber\\ &\mbox{s.t. }
\|(\bm{\mathcal{R}}_{i}\bfm) \odot (\bm{\mathcal{R}}_{i}\hat{\mathbf{s}} - \mathbf{\Phi Ax})\|_2^2 \leq \|\bm{\mathcal{R}}_{i}\bfm\|^{\phantom{,}}_0 \, \sigma^2,\quad\forall i = 1, \dots, R,
\end{eqnarray}
where the mask $\bfm$ guarantees that the missing traces are not taken into account and $\sigma^2$ here is the variance of the noise assuming additive white Gaussian noise.

Then, in the process of updating each column $\mathbf{a}_k$ of the matrix $\mathbf{A}$ using the fixed $\hat{\mathbf{s}}$ and calculated $\mathbf{x}_{i}$, the following problem, which replaces (\ref{eq:sparse-ksvd-rank1-approximation}), needs to be solved:
\begin{equation}
\label{eq:inpainting-rank1-approximation}
\left\{\mathbf{a}_k, \mathbf{X}_{k,\mathcal{I}_k}^T\right\} = \argmin\limits_{\mathbf{a}, \mathbf{x}} \|\mathbf{M}_k \odot (\mathbf{E}_k - \mathbf{\Phi ax}^T)\|_F^2
\quad \mbox{s.t. }
\begin{cases}
	\|\mathbf{a}\|_1 \leq p\\
	\|\mathbf{\Phi a}\|_2 = 1,
	\end{cases}
\end{equation}
where the matrix $\mathbf{M}_k$ collects $\bm{\mathcal{R}}_{i}\bfm$ in columns for those $i \in \mathcal{I}_k$ and it has the same size as $\mathbf{E}_k$. Different from (\ref{eq:sparse-ksvd-rank1-approximation}), this problem is a weighted low-rank approximation problem. Unfortunately, due to the element-wise mask matrix $\mathbf{M}_k$, we cannot explicitly find the simple form by Lemma 1 in \cite{Rubinstein:2010aa} as we did for (\ref{eq:sparse-ksvd-rank1-approximation}) and (\ref{eq:simpleform}).
Alternatively, Nati and Jaakkola \cite{Nati:2003aa} put forward a simple but effective iterative algorithm that converges to the local minima of the objective function in (\ref{eq:inpainting-rank1-approximation}). The algorithm is based on the expectation-maximization (EM) procedure in which the expectation step fills in the current estimate of $\mathbf{\Phi ax}^T$ for all missing elements in $\mathbf{M}_k \odot \mathbf{E}_k$ and the maximization step updates $\mathbf{\Phi ax}^T$ from the filled-in version of $\mathbf{M}_k \odot \mathbf{E}_k$.
\begin{algorithm}
	\caption{Weighted low-rank approximation algorithm}
	\label{alg:weighted-sparse-ksvd-rank1-approximation}
	% \SetAlgoLined
	\SetAlFnt{\small}
	\AlFnt
	\SetKwInOut{KwInit}{Initialization}
	\KwIn{$\mathbf{E}_k \in \mathbb{R}^{N \times |\mathcal{I}_k|}$, base dictionary $\mathbf{\Phi} \in \mathbb{R}^{N \times L}$, mask matrix $\mathbf{M}_k \in \mathbb{R}^{N \times |\mathcal{I}_k|}$, \rule{10mm}{0mm} {number of iterations $K$\rule[-1.2mm]{0mm}{4mm}}}
	\KwOut{$\mathbf{a}_k \in \mathbb{R}^L$, $\mathbf{X}_{k,\mathcal{I}_k}^T \in \mathbb{R}^{|\mathcal{I}_k|}$}
	\KwInit{$\mathbf{a}_{\mbox{new}} \gets \mathbf{0}$, $\mathbf{x}_{\mbox{new}} \gets \mathbf{X}_{k, \mathcal{I}_k}^T$}
	\Repeat{$K$ iterations}
	
		$\mathbf{a}_{\mbox{old}} \gets \mathbf{a}_{\mbox{new}}$\\
		$\mathbf{x}_{\mbox{old}} \gets \mathbf{x}_{\mbox{new}}$\\
		Solve the following problem with the assistance of Lemma 1 in \cite{Rubinstein:2010aa}
		$\{\mathbf{a}_{\mbox{new}}, \mathbf{x}_{\mbox{new}}\} = \begin{cases}
			\argmin\limits_{\mathbf{a}, \mathbf{x}} \biggl\| \overbrace{\left[ \mathbf{M}_k \odot \mathbf{E}_k + (\mathbf{1} - \mathbf{M}_k) \odot (\mathbf{\Phi}\mathbf{a}^{\phantom{,}}_{\mbox{old}}\mathbf{x}_{\mbox{old}}^{T\phantom{,}}) \right]}^{\mathbf{E}_k'} - \mathbf{\Phi ax}^T \biggr\|_F^2\\
			\mbox{subject to} \;\; \|\mathbf{a}\|_1 \leq p \mbox{\ \,and\,\ } \|\mathbf{\Phi a}\|_2 = 1\\
			\end{cases}$\\
		$\mathbf{a}^{\phantom{,}}_k \gets \mathbf{a}^{\phantom{,}}_{\mbox{new}}$\\
		$\mathbf{X}_{k,\mathcal{I}_k}^T \gets \mathbf{x}^{\phantom{,}}_{\mbox{new}}$\\
	
\end{algorithm}

Concretely, Algorithm \ref{alg:weighted-sparse-ksvd-rank1-approximation} presents the iterative EM-based algorithm that solves (\ref{eq:inpainting-rank1-approximation}). Every time $\mathbf{a}$ and $\mathbf{d}$ are estimated, with the names $\mathbf{a}_{\mbox{old}}$ and $\mathbf{x}_{\mbox{old}}$, they are used to fill in $\mathbf{M}_k \odot \mathbf{E}_k$ by generating a new observation matrix
\begin{equation}
\mathbf{E}_k' \triangleq \mathbf{M}_k \odot \mathbf{E}_k + (\mathbf{1} - \mathbf{M}_k) \odot (\mathbf{\Phi}\mathbf{a}_{\mbox{old}}\mathbf{x}_{\mbox{old}}^T)
\end{equation}
in the expectation step. Then, in the maximization step, $\mathbf{a}$ and $\mathbf{d}$ are updated by the filled-in observation matrix $\mathbf{E}_k'$
\begin{equation}
\label{eq:inpainting-rank1-approximation-iteration}
\{\mathbf{a}_{\mbox{new}}, \mathbf{x}_{\mbox{new}}\} = \argmin\limits_{\mathbf{a}, \mathbf{x}} \left\| \mathbf{E}_k' - \mathbf{\Phi ax}^T \right\|_F^2 
\quad \mbox{s.t. }
\begin{cases}
	\|\mathbf{a}\|_1 \leq p\\
	\|\mathbf{\Phi a}\|_2 = 1.\\
	\end{cases}
\end{equation}
The problem in the form of (\ref{eq:inpainting-rank1-approximation-iteration}) can be solved with using Lemma 1 in \cite{Rubinstein:2010aa} as in sparse K-SVD. The EM procedure converges to a local minimum very quickly, within only a few ($K \approx 5$) iterations.

Finally, with $\mathbf{A}$ and all $\mathbf{x}_{i}$ obtained, the last remaining problem of (\ref{eq:fused-dataset-learning-inpainting}) for the interpolation result $\hat{\mathbf{s}}$ is the least-squares problem
\begin{equation}
\label{eq:dataset-inpainting-least-squares}
\hat{\mathbf{s}} = \argmin\limits_{\mathbf{s}} \sum\limits_{i} \|\bm{\mathcal{R}}_{i}\mathbf{s} - \mathbf{\Phi Ax}_{i}\|_2^2 + \lambda\|\bfm \odot (\mathbf{s} - \mathbf{w})\|_2^2.
\end{equation}
which has the closed-form solution 
\begin{equation}
\label{eq:dataset-inpainting-least-squares-solution}
\hat{\mathbf{s}} = \left( \lambda\mbox{diag}(\bfm) + \sum_i \bm{\mathcal{R}}_i^{\dag}\bm{\mathcal{R}}_i \right)^{-1} \left( \lambda(\bfm \odot \mathbf{w}) + \sum\limits_{i}\bm{\mathcal{R}}_i^{\dag}\mathbf{\Phi A}\mathbf{x}_i \right).
\end{equation}
Note that the mask $\bm{\mathcal{R}}_{i}\bfm$ has been removed in front of the reconstruction misfit $\bm{\mathcal{R}}_{i}\mathbf{s} - \mathbf{\Phi Ax}_{i}$ in (\ref{eq:dataset-inpainting-least-squares}) since at this point the entire $\mathbf{s}$ is now being restored including the missing traces.

The detailed implementation of the proposed seismic data recovery method can be found in Algorithm \ref{alg:dataset-inpainting}, where the atom replacing is a trick borrowed from \cite{Zhu:2015aa} which replaces those duplicated and rarely used atoms in the learned dictionary and in turn improves the efficiency of the dictionary learning.

\begin{algorithm}
	\caption{Recover seismic dataset using the double-sparsity dictionary learned on patches from the noisy dataset with missing traces}
	\label{alg:dataset-inpainting}
	% \SetAlgoLined
	\SetAlFnt{\small}
	\AlFnt
	\SetKwInOut{KwInit}{Initialization}
	\KwIn{Vectorized noisy seismic dataset $\mathbf{w} \in \mathbb{R}^{N_zN_x}$ with missing traces, 
		mask vector $\bfm \in \mathbb{R}^{N_zN_x}$ patch height $n_z$, patch width $n_x$, $N = n_zn_x$,  
		base dictionary $\mathbf{\Phi} \in \mathbb{R}^{N \times L}$, number of training iterations $K_T$, 
		number of atom update iterations $K_U$}
	\KwOut{Interpolated seismic dataset $\hat{\mathbf{s}} \in \mathbb{R}^{N_zN_x}$, sparse matrix $\mathbf{A} \in \mathbb{R}^{L \times L}$,   
		sparse coefficient matrix $\mathbf{X} \in \mathbb{R}^{L \times R}$}
	\KwInit{ $\hat{\mathbf{s}} \gets \mathbf{w}$, $\mathbf{A} \gets \mathbf{I}$, $\mathbf{X} \gets \mathbf{0}$ }
	\Repeat{$K_T$ training iterations}
	{
		\tcp{Sparse Representation Stage}
		\For{$i \gets 1$ \KwTo $R$}
		{
			
			$\mathbf{x}_{i} \gets \argmin\limits_{\mathbf{x}} \|\mathbf{x}\|_1\ \mbox{\,\ s.t.}\ \|(\bm{\mathcal{R}}_{i}\bfm) \odot (\bm{\mathcal{R}}_{i}\hat{\mathbf{s}} - \mathbf{\Phi Ax})\|_2^2 \leq \|\bm{\mathcal{R}}_{i}\bfm\|^{\phantom{,}}_0 \, \sigma^2$\\
			Place $\mathbf{x}_{i}$ into $\mathbf{X}$ as a column at the corresponding position;
			
		}
		\tcp{Dictionary Update Stage}
		\For{$k \gets 1$ \KwTo $L$}
		{
			$\mathcal{I}_k \gets \{ r | 1 \leq r \leq R, x_{kr} \neq 0 \}$\\
			$\mathbf{M}_k$ collects $\bm{\mathcal{R}}_{i}\bfm$ in columns for those $i$ that satisfy $i \in \mathcal{I}_k$\\
			\tcp{Atom removal}
			$\mathbf{a}_{\mbox{new}} \gets \mathbf{a}_k \gets \mathbf{0}$\\
			$\mathbf{x}_{\mbox{new}} \gets \mathbf{X}_{k, \mathcal{I}_k}^T$\\
			$\mathbf{E}_k \gets \mathbf{Y}_{\mathcal{I}_k} - \mathbf{\Phi AX}_{\mathcal{I}_k}$\\
			\tcp{Atom updating}
			Use weighted low-rank approximation (Algorithm \ref{alg:weighted-sparse-ksvd-rank1-approximation}) to find
			$\mathbf{a}_k $ and $\mathbf{X}_{k, \mathcal{I}_k}$\\
		}
		\For{$k \gets 1$ \KwTo $L$}
		{
			Atom\_Replacing($\mathbf{\Phi a}_k$)\\
		}
	}
	\tcp{Interpolation Stage}
	$\hat{\mathbf{s}} \gets \left( \lambda\mbox{diag}(\bfm) + \sum\limits_i \bm{\mathcal{R}}_{i}^{\dag} \bm{\mathcal{R}}_{i} \right)^{-1} \left( \lambda(\bfm\odot \mathbf{w}) + \sum\limits_{i} \bm{\mathcal{R}}_{i}^{\dag}\mathbf{\Phi A}\mathbf{x}_{i} \right)$
\end{algorithm}

Concretely, Algorithm \ref{alg:weighted-sparse-ksvd-rank1-approximation} presents the iterative EM-based algorithm that solves (\ref{eq:inpainting-rank1-approximation}). Every time $\mathbf{a}$ and $\mathbf{d}$ are estimated, with the names $\mathbf{a}_{\mbox{old}}$ and $\mathbf{x}_{\mbox{old}}$, they are used to fill in $\mathbf{M}_k \odot \mathbf{E}_k$ by generating a new observation matrix
\begin{equation}
\mathbf{E}_k' \triangleq \mathbf{M}_k \odot \mathbf{E}_k + (\mathbf{1} - \mathbf{M}_k) \odot (\mathbf{\Phi}\mathbf{a}_{\mbox{old}}\mathbf{x}_{\mbox{old}}^T)
\end{equation}
in the expectation step. Then, in the maximization step, $\mathbf{a}$ and $\mathbf{d}$ are updated by the filled-in observation matrix $\mathbf{E}_k'$
\begin{equation}
\label{eq:inpainting-rank1-approximation-iteration}
\{\mathbf{a}_{\mbox{new}}, \mathbf{x}_{\mbox{new}}\} = \argmin\limits_{\mathbf{a}, \mathbf{x}} \left\| \mathbf{E}_k' - \mathbf{\Phi ax}^T \right\|_F^2 
\quad \mbox{s.t. }
\begin{cases}
	\|\mathbf{a}\|_1 \leq p\\
	\|\mathbf{\Phi a}\|_2 = 1.
	\end{cases}
\end{equation}
The problem in the form of (\ref{eq:inpainting-rank1-approximation-iteration}) can be solved with using Lemma 1 in \cite{Rubinstein:2010aa} as in sparse K-SVD. The EM procedure converges to a local minimum very quickly, within only a few ($K \approx 5$) iterations.

Finally, with $\mathbf{A}$ and all $\mathbf{x}_{i}$ obtained, the last remaining problem of (\ref{eq:fused-dataset-learning-inpainting}) for the interpolation result $\hat{\mathbf{s}}$ is the least-squares problem
\begin{equation}
\label{eq:dataset-inpainting-least-squares}
\hat{\mathbf{s}} = \argmin\limits_{\mathbf{s}} \sum\limits_{i} \|\bm{\mathcal{R}}_{i}\mathbf{s} - \mathbf{\Phi Ax}_{i}\|_2^2 + \lambda\|\bfm \odot (\mathbf{s} - \mathbf{w})\|_2^2.
\end{equation}
which has the closed-form solution 
\begin{equation}
\label{eq:dataset-inpainting-least-squares-solution}
\hat{\mathbf{s}} = \left( \lambda\mbox{diag}(\bfm) + \sum_i \bm{\mathcal{R}}_i^{\dag}\bm{\mathcal{R}}_i \right)^{-1} \left( \lambda(\bfm \odot \mathbf{w}) + \sum\limits_{i}\bm{\mathcal{R}}_i^{\dag}\mathbf{\Phi A}\mathbf{x}_i \right).
\end{equation}
Note that the mask $\bm{\mathcal{R}}_{i}\bfm$ has been removed in front of the reconstruction misfit $\bm{\mathcal{R}}_{i}\mathbf{s} - \mathbf{\Phi Ax}_{i}$ in (\ref{eq:dataset-inpainting-least-squares}) since at this point the entire $\mathbf{s}$ is now being restored including the missing traces.

The detailed implementation of the proposed seismic data recovery method can be found in Algorithm \ref{alg:dataset-inpainting}, where the atom replacing is a trick borrowed from \cite{Zhu:2015aa} which replaces those duplicated and rarely used atoms in the learned dictionary and in turn improves the efficiency of the dictionary learning.

\begin{algorithm}
	\caption{Recover seismic dataset using the double-sparsity dictionary learned on patches from the noisy dataset with missing traces}
	\label{alg:dataset-inpainting}
	% \SetAlgoLined
	\SetAlFnt{\small}
	\AlFnt
	\SetKwInOut{KwInit}{Initialization}
	\KwIn{Vectorized noisy seismic dataset $\mathbf{w} \in \mathbb{R}^{N_zN_x}$ with missing traces, 
		mask vector $\bfm \in \mathbb{R}^{N_zN_x}$ patch height $n_z$, patch width $n_x$, $N = n_zn_x$,  
		base dictionary $\mathbf{\Phi} \in \mathbb{R}^{N \times L}$, number of training iterations $K_T$, 
		number of atom update iterations $K_U$}
	\KwOut{Interpolated seismic dataset $\hat{\mathbf{s}} \in \mathbb{R}^{N_zN_x}$, sparse matrix $\mathbf{A} \in \mathbb{R}^{L \times L}$,   
		sparse coefficient matrix $\mathbf{X} \in \mathbb{R}^{L \times R}$}
	\KwInit{ $\hat{\mathbf{s}} \gets \mathbf{w}$, $\mathbf{A} \gets \mathbf{I}$, $\mathbf{X} \gets \mathbf{0}$ }
	\Repeat{$K_T$ training iterations}
	{
		\tcp{Sparse Representation Stage}
		\For{$i \gets 1$ \KwTo $R$}
		{
			
			$\mathbf{x}_{i} \gets \argmin\limits_{\mathbf{x}} \|\mathbf{x}\|_1\ \mbox{\,\ s.t.}\ \|(\bm{\mathcal{R}}_{i}\bfm) \odot (\bm{\mathcal{R}}_{i}\hat{\mathbf{s}} - \mathbf{\Phi Ax})\|_2^2 \leq \|\bm{\mathcal{R}}_{i}\bfm\|^{\phantom{,}}_0 \, \sigma^2$\\
			Place $\mathbf{x}_{i}$ into $\mathbf{X}$ as a column at the corresponding position;
			
		}
		\tcp{Dictionary Update Stage}
		\For{$k \gets 1$ \KwTo $L$}
		{
			$\mathcal{I}_k \gets \{ r | 1 \leq r \leq R, x_{kr} \neq 0 \}$\\
			$\mathbf{M}_k$ collects $\bm{\mathcal{R}}_{i}\bfm$ in columns for those $i$ that satisfy $i \in \mathcal{I}_k$\\
			\tcp{Atom removal}
			$\mathbf{a}_{\mbox{new}} \gets \mathbf{a}_k \gets \mathbf{0}$\\
			$\mathbf{x}_{\mbox{new}} \gets \mathbf{X}_{k, \mathcal{I}_k}^T$\\
			$\mathbf{E}_k \gets \mathbf{Y}_{\mathcal{I}_k} - \mathbf{\Phi AX}_{\mathcal{I}_k}$\\
			\tcp{Atom updating}
			Use weighted low-rank approximation (Algorithm \ref{alg:weighted-sparse-ksvd-rank1-approximation}) to find
			$\mathbf{a}_k $ and $\mathbf{X}_{k, \mathcal{I}_k}$\\
		}
		\For{$k \gets 1$ \KwTo $L$}
		{
			Atom\_Replacing($\mathbf{\Phi a}_k$)\\
		}
	}
	\tcp{Interpolation Stage}
	$\hat{\mathbf{s}} \gets \left( \lambda\mbox{diag}(\bfm) + \sum\limits_i \bm{\mathcal{R}}_{i}^{\dag} \bm{\mathcal{R}}_{i} \right)^{-1} \left( \lambda(\bfm\odot \mathbf{w}) + \sum\limits_{i} \bm{\mathcal{R}}_{i}^{\dag}\mathbf{\Phi A}\mathbf{x}_{i} \right)$
\end{algorithm}

\section{Numerical Simulation}\label{sec:numerical}
The following experiments provide recovery performance results for the double-sparsity dictionary learning method when the seismic dataset has missing traces and is corrupted with additive random noise. Figure \ref{fig:bppublic_data_for_inpainting}(a) shows the original dataset provided by BP \cite{Etgen:1998aa,Madagascar} where the number of receivers is $N_x = 240$ and each trace has $N_z = 384$ time samples. Noisy seismic datasets are shown in Figure \ref{fig:bppublic_data_for_inpainting}(b,c) with 33\% and 50\% missing traces whose indices are randomly selected between 1 and 240. Note that all the missing traces have Not-a-Number (NaN) values and their corresponding values in the mask vector $\bfm$ are set to zeros. For the valid (non-missing) traces, white Gaussian noise with $\sigma = 0.1$ is added. The value of $\sigma$ is chosen after we remove the mean value of each trace and then normalize its range to one, i.e., divide by the difference between max and min values. After this normalization the largest absolute value on a valid trace is approximately 0.5.

\begin{figure}[h!]
	\centering
	\subfloat[Contourlet BPDN for inpainting (PSNR = 27.50\,dB)][Contourlet BPDN for inpainting \\  (PSNR = 27.50\,dB)]
	{
		\begin{minipage}{0.4\linewidth}
			\centering
			\includegraphics[width=\textwidth]{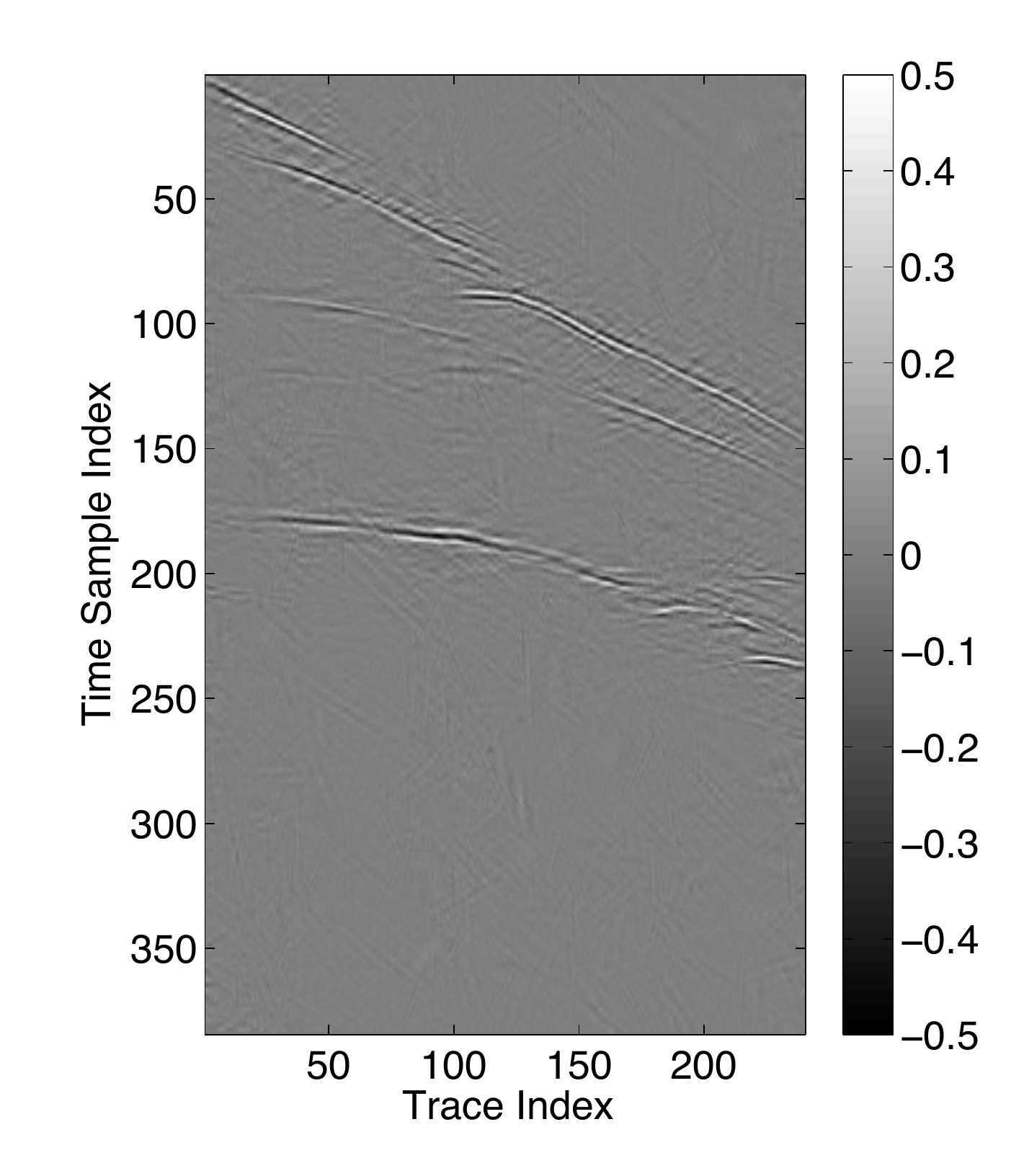}
		\end{minipage}
		\label{fig:bppublic_inpaintedData_contourlet_bpdn_033}
	}
	\subfloat[Error Panel for Counterlet]
	{
		\begin{minipage}{0.4\linewidth}
			\centering
			\includegraphics[width=\textwidth]{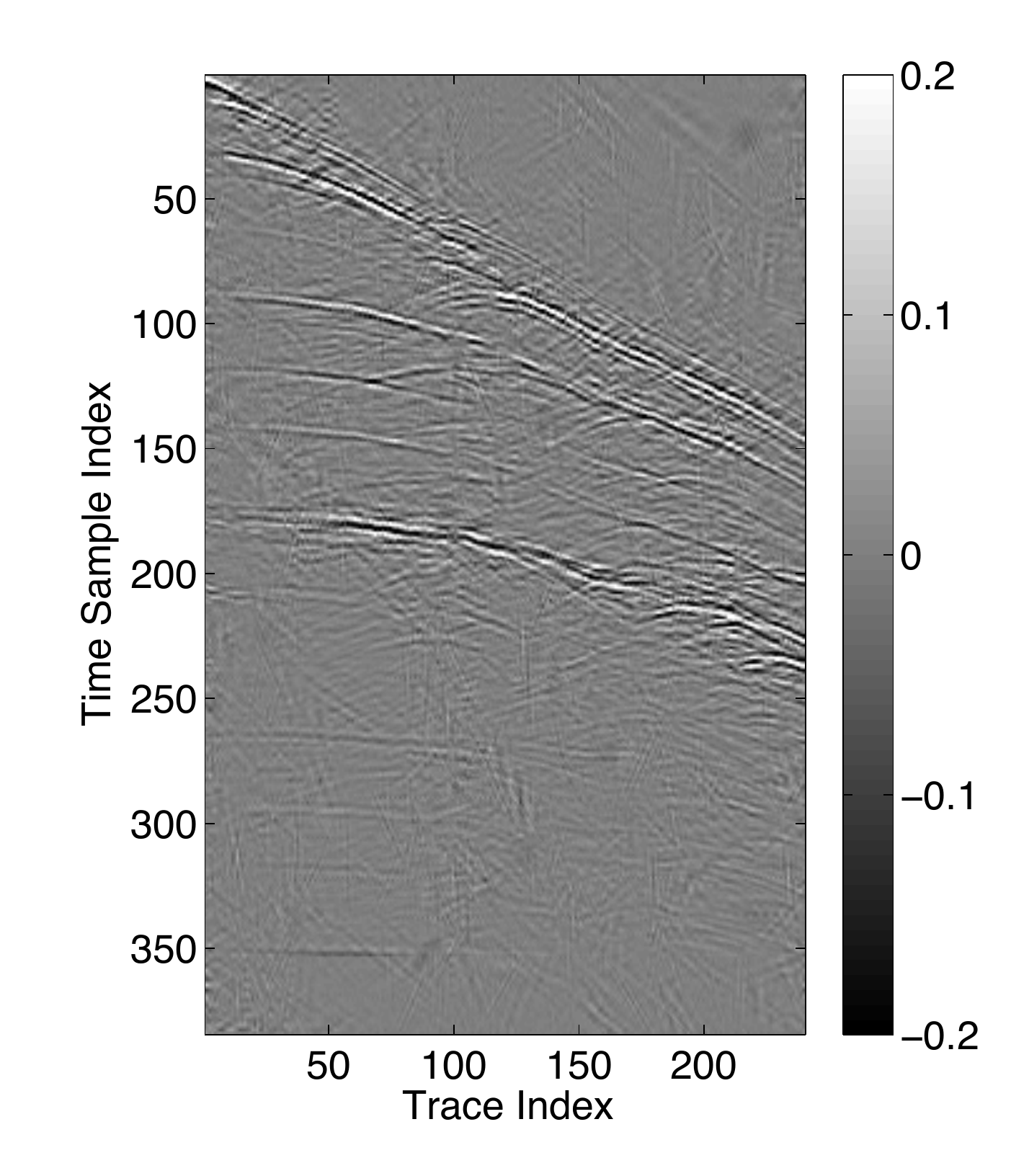}
		\end{minipage}
		\label{fig:bppublic_inpaintedDiffData_contourlet_bpdn}
	}\\
	\subfloat[Curvelet BPDN for inpainting (PSNR = 28.12\,dB)][Curvelet BPDN for inpainting \\ (PSNR = 28.12\,dB)]
	{
		\begin{minipage}{0.4\linewidth}
			\centering
			\includegraphics[width=\textwidth]{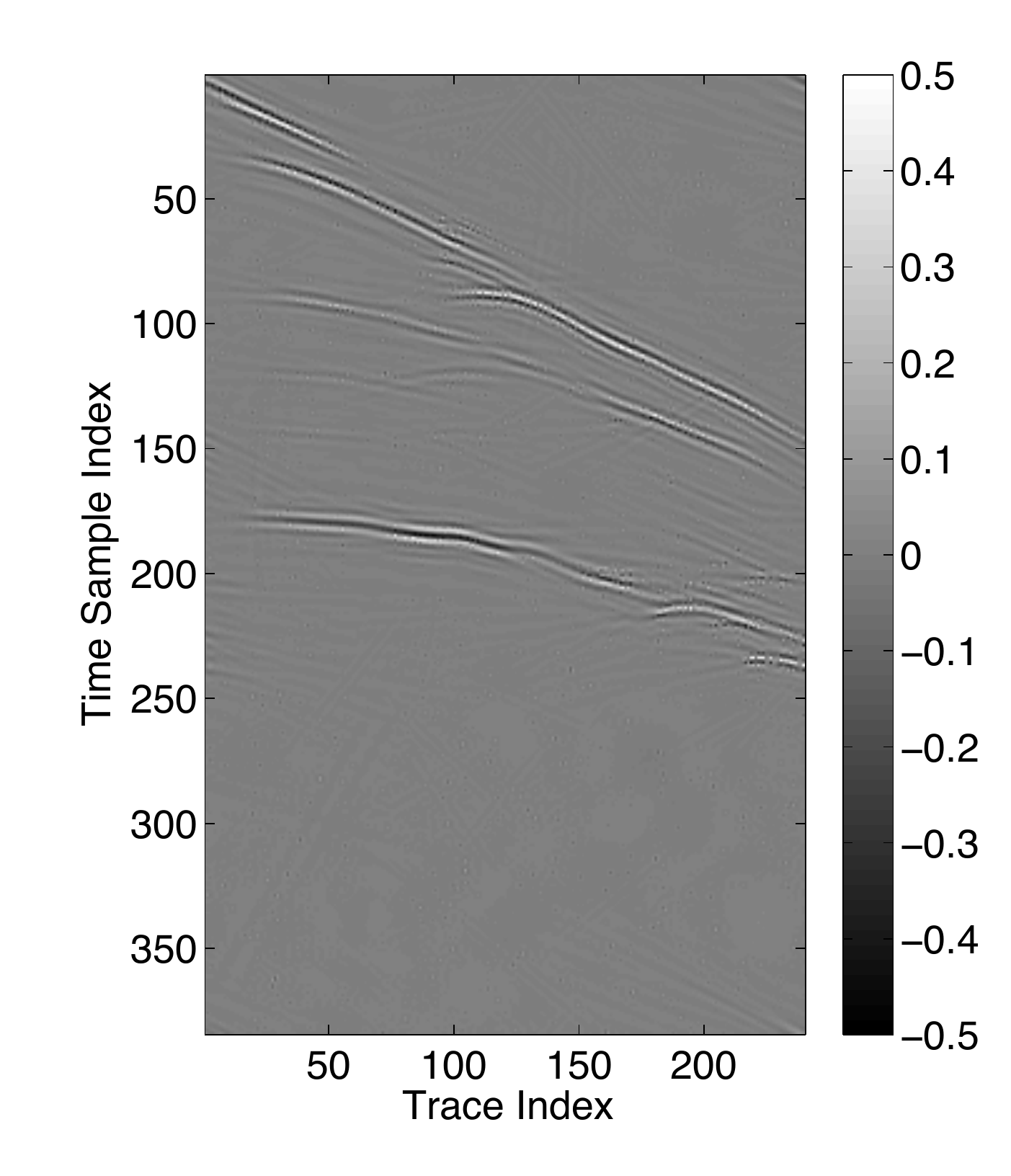}
		\end{minipage}
		\label{fig:bppublic_inpaintedData_curvelet_bpdn}
	}
	\subfloat[Error Panel for Curvelet]
	{
		\begin{minipage}{0.4\linewidth}
			\centering
			\includegraphics[width=\textwidth]{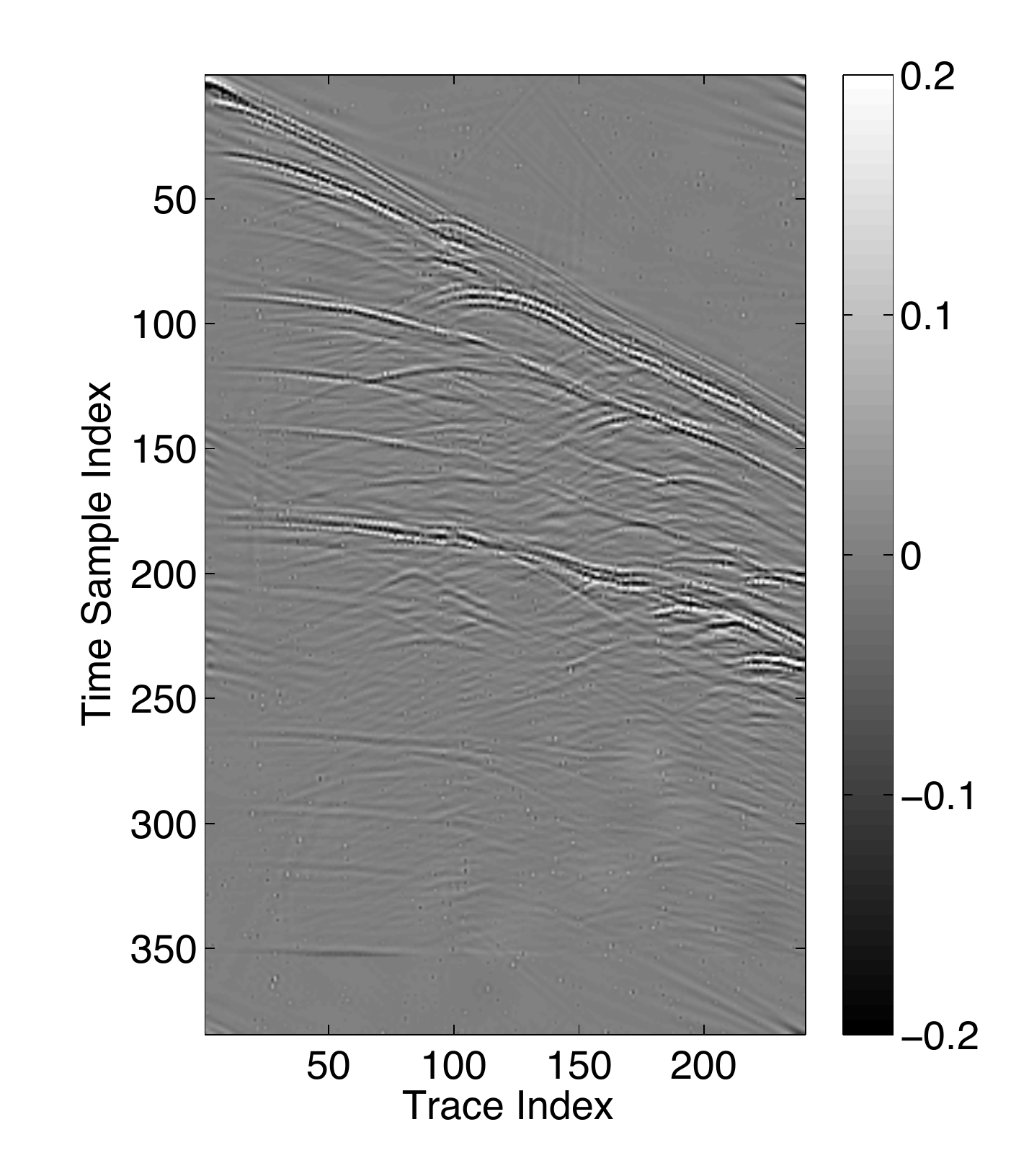}
		\end{minipage}
		\label{fig:bppublic_inpaintedDiffData_curvelet_bpdn}
	}
	\caption{Recovery results for 33\% missing traces based on BPDN using the fixed multi-scale transforms: (a) result by contourlet-based BPDN method (PSNR = 27.50\,dB), (b) is the difference between (a) and the original data, (c) result by curvelet-based BPDN method (PSNR = 28.12\,dB), and (d) is the difference between (c) and the original data. Note the change of gray scale for the error panels.}
	\label{fig:bppublic_inpainting_transforms}
\end{figure}

First, as baseline experiments, the fixed multi-scale contourlet and curvelet transforms are used for seismic dataset recovery (denoising and interpolation jointly). The BPDN method (implemented by the package SPGL1 \cite{BergFriedlander:2008}) is used to find the sparse representation of all the valid traces and then the missing traces are inferred via inverse transform operations as follows
\begin{equation}
\label{eq:inpainting-fix-transforms-bpdn}
\begin{cases}
	\hat{\mathbf{x}} = \argmin\limits_{\mathbf{x}} \|\mathbf{x}\|_1 \quad \mbox{s.t.\ \;} \|\bfm \odot (\mathbf{w} - \mathbf{\Phi x})\|_2^2 \leq \|\bfm\|^{\phantom{,}}_0 \, \sigma^2\\
	\hat{\mathbf{s}} = \mathbf{\Phi}\hat{\mathbf{x}}
	\end{cases}
\end{equation}
where $\mathbf{\Phi}$ refers to the dictionary of the contourlet/curvelet synthesis operator. Figure \ref{fig:bppublic_inpainting_transforms} presents the restoration results based on the BPDN method using the contourlet and curvelet transforms for the 33\% missing traces case. The performance using  contourlets can achieve PSNR = 27.50\,dB while using the curvelets can achieve PSNR = 28.12\,dB. Still, just like in the pure denoising scenario \cite{Zhu:2015aa}, pseudo-Gibbs artifacts are quite obvious in the recovery results.

\begin{figure}[h!]
	\centering
	\subfloat[DCT Dictionary $\mathbf{\Phi}$]
	{
		\begin{minipage}{0.4\linewidth}
			\centering
			\includegraphics[width=\textwidth]{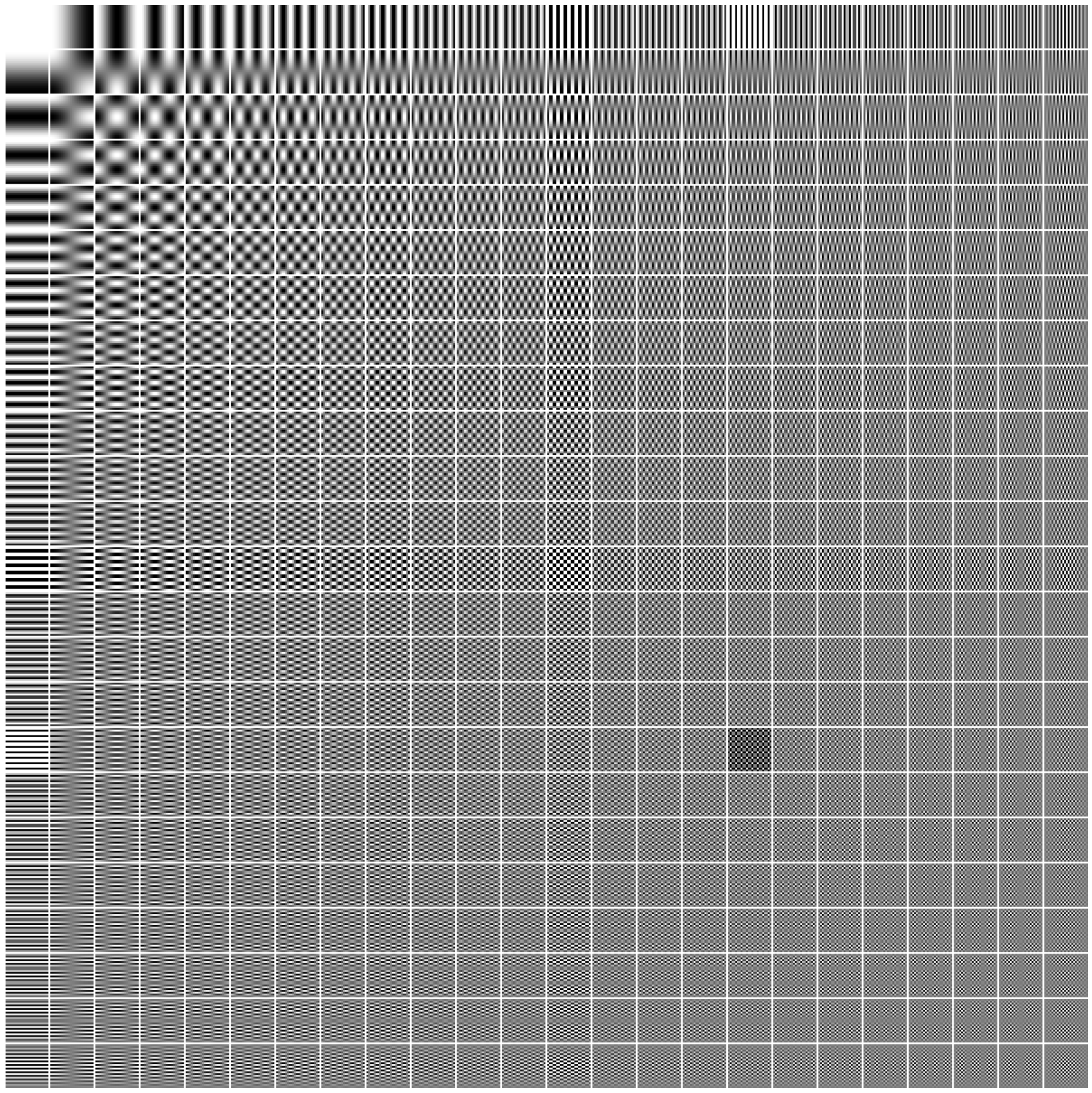}
		\end{minipage}
		\label{fig:dctAtoms_576x576}
	}
	\subfloat[Learned Matrix $\mathbf{A}$]
	{
		\begin{minipage}{0.4\linewidth}
			\centering
			\includegraphics[width=\textwidth]{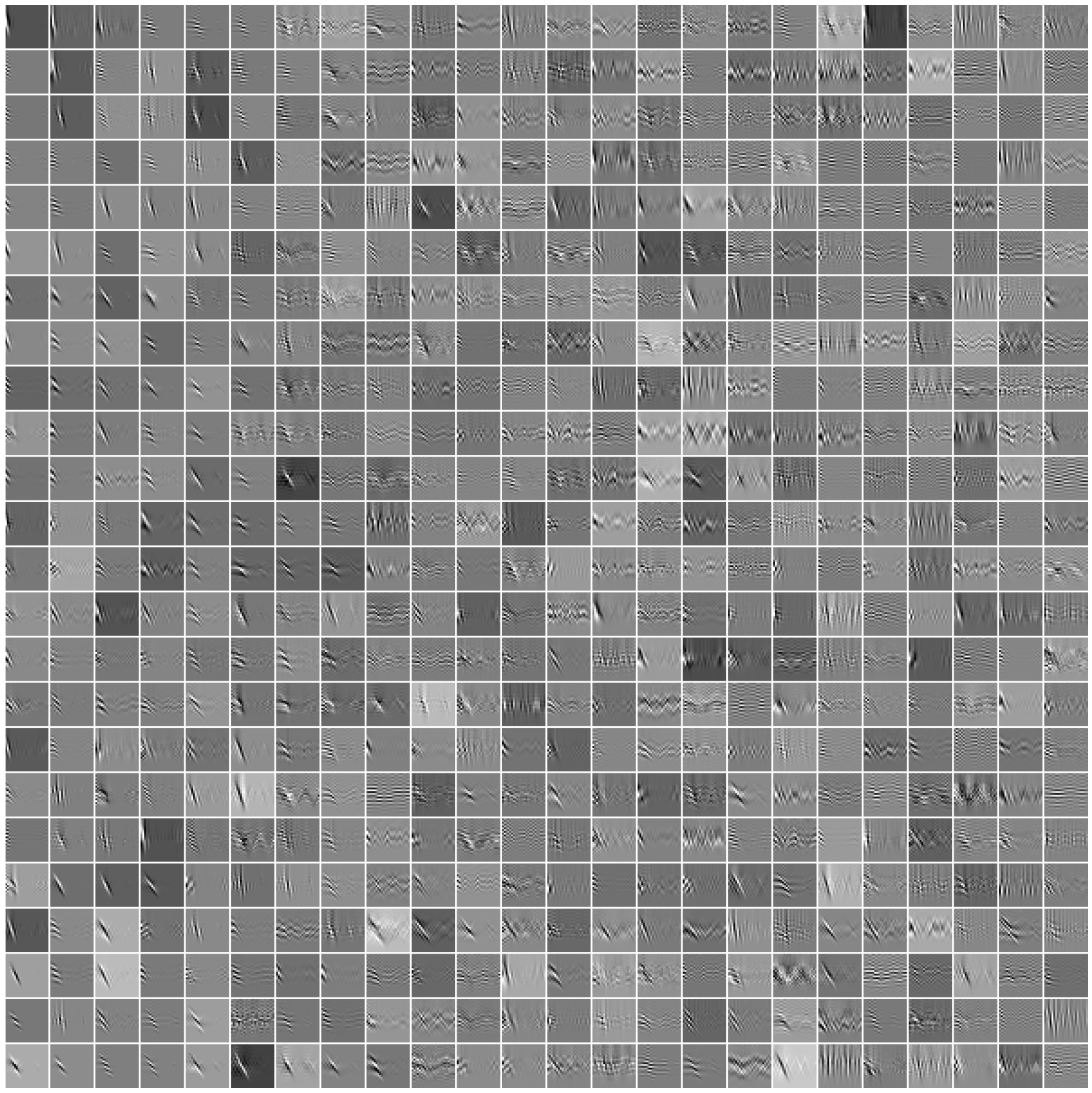}
		\end{minipage}
		\label{fig:bppublic_A_dct_forInpainting}
	}\\
	\subfloat[Overall Learned Dictionary $\mathbf{D} = \mathbf{\Phi A}$]
	{
		\begin{minipage}{0.4\linewidth}
			\centering
			\includegraphics[width=\textwidth]{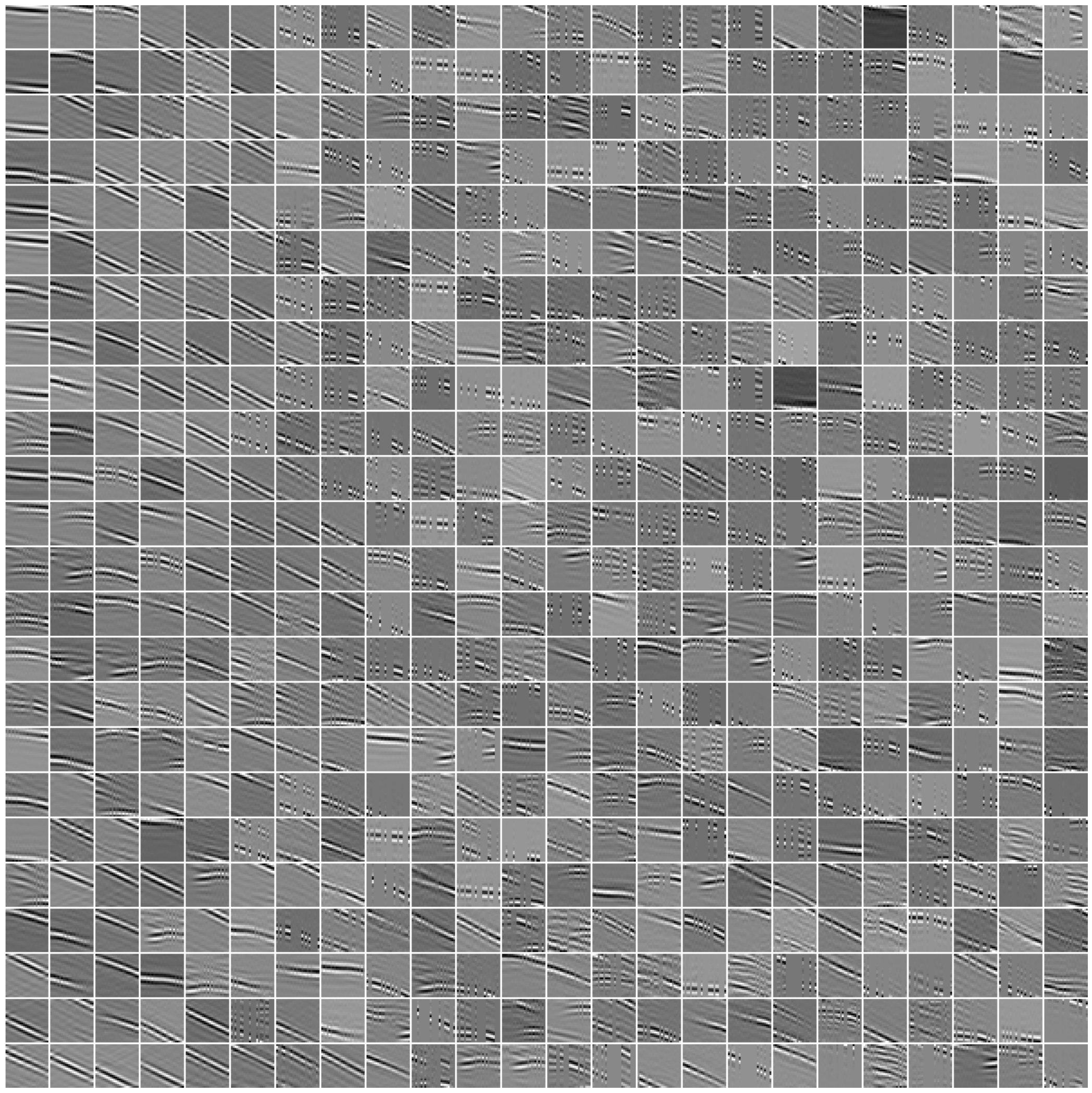}
		\end{minipage}
		\label{fig:bppublic_D_dct_forInpainting}
	}
	\caption{Base dictionary (DCT) and learned dictionaries.
		Individal patches are $24\times24$, and the dictionaries contain 576 patches arranged as a $24\times24$ grid.}
	\label{fig:bppublic_denoising_sparseKsvD_dict}
\end{figure}
\begin{figure}[h!]
	\centering
	\subfloat[Result by $\mathbf{D} = \mathbf{\Phi A}$ (PSNR = 32.11\,dB)][Result by $\mathbf{D} = \mathbf{\Phi A}$ , 33\% missing\\ (PSNR = 32.11\,dB)]
	{
		\begin{minipage}{0.4\linewidth}
			\centering
			\includegraphics[width=\textwidth]{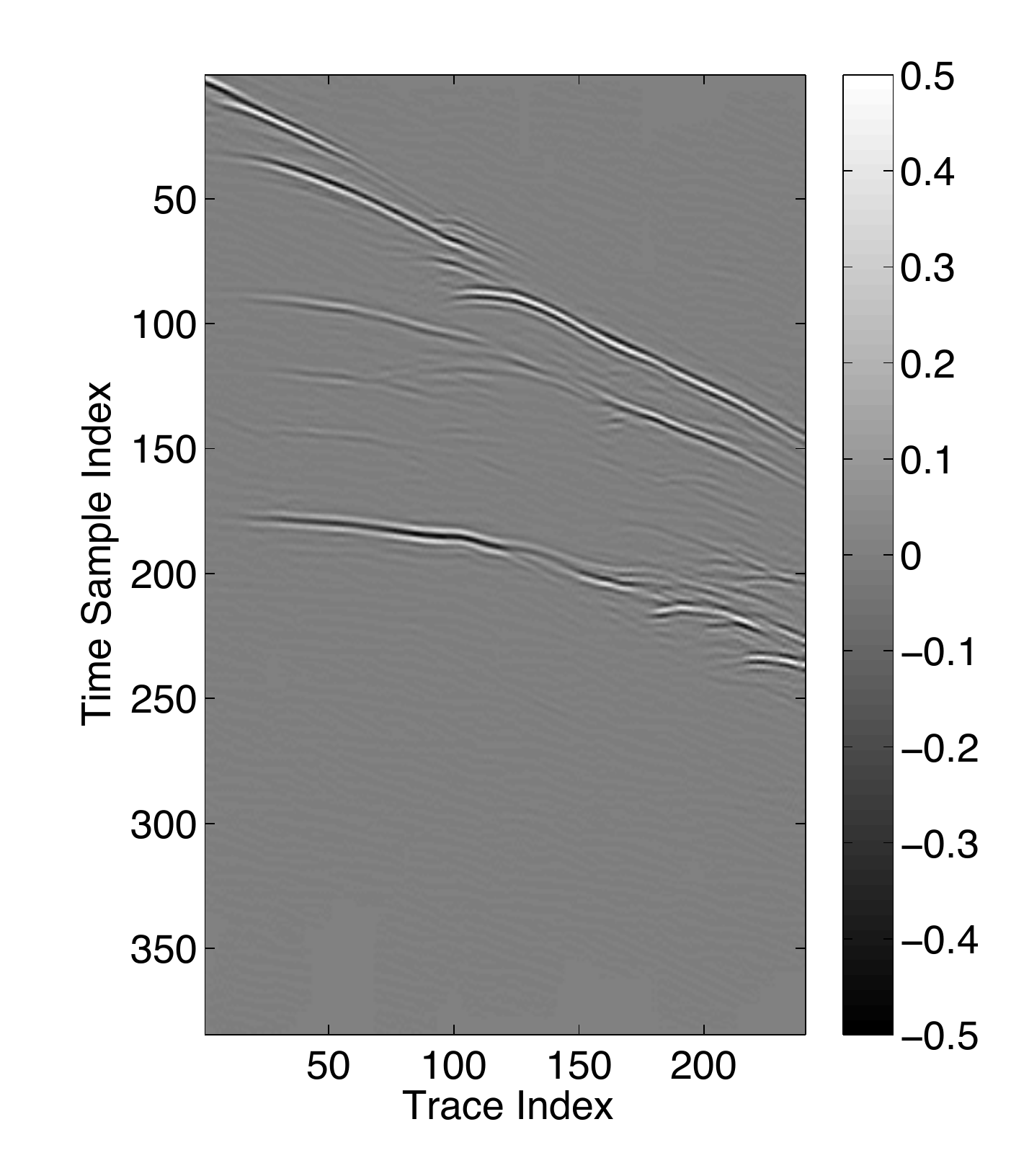}
		\end{minipage}
		\label{fig:bppublic_inpaintedData_sparseKsvd_dct_033}
	}
	\subfloat[Error Panel, 33\% missing]
	{
		\begin{minipage}{0.4\linewidth}
			\centering
			\includegraphics[width=\textwidth]{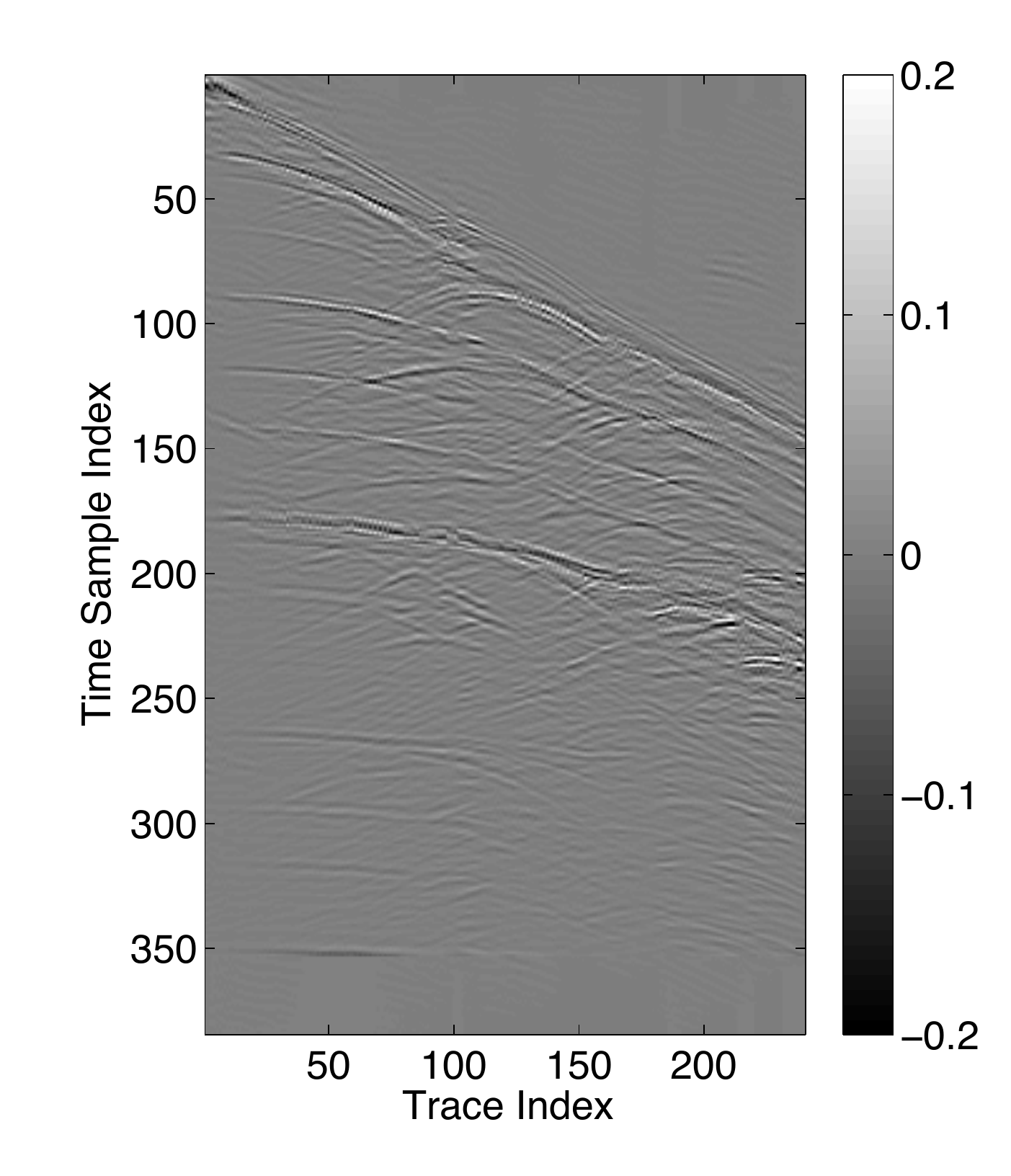}
		\end{minipage}
		\label{fig:bppublic_inpaintedDiffData_sparseKsvd_dct_033}
	}\\
	\subfloat[Result by $\mathbf{D} = \mathbf{\Phi A}$ (PSNR = 30.31\,dB)][Result by $\mathbf{D} = \mathbf{\Phi A}$, 50\% missing \\ (PSNR = 30.31\,dB)]
	{
		\begin{minipage}{0.4\linewidth}
			\centering
			\includegraphics[width=\textwidth]{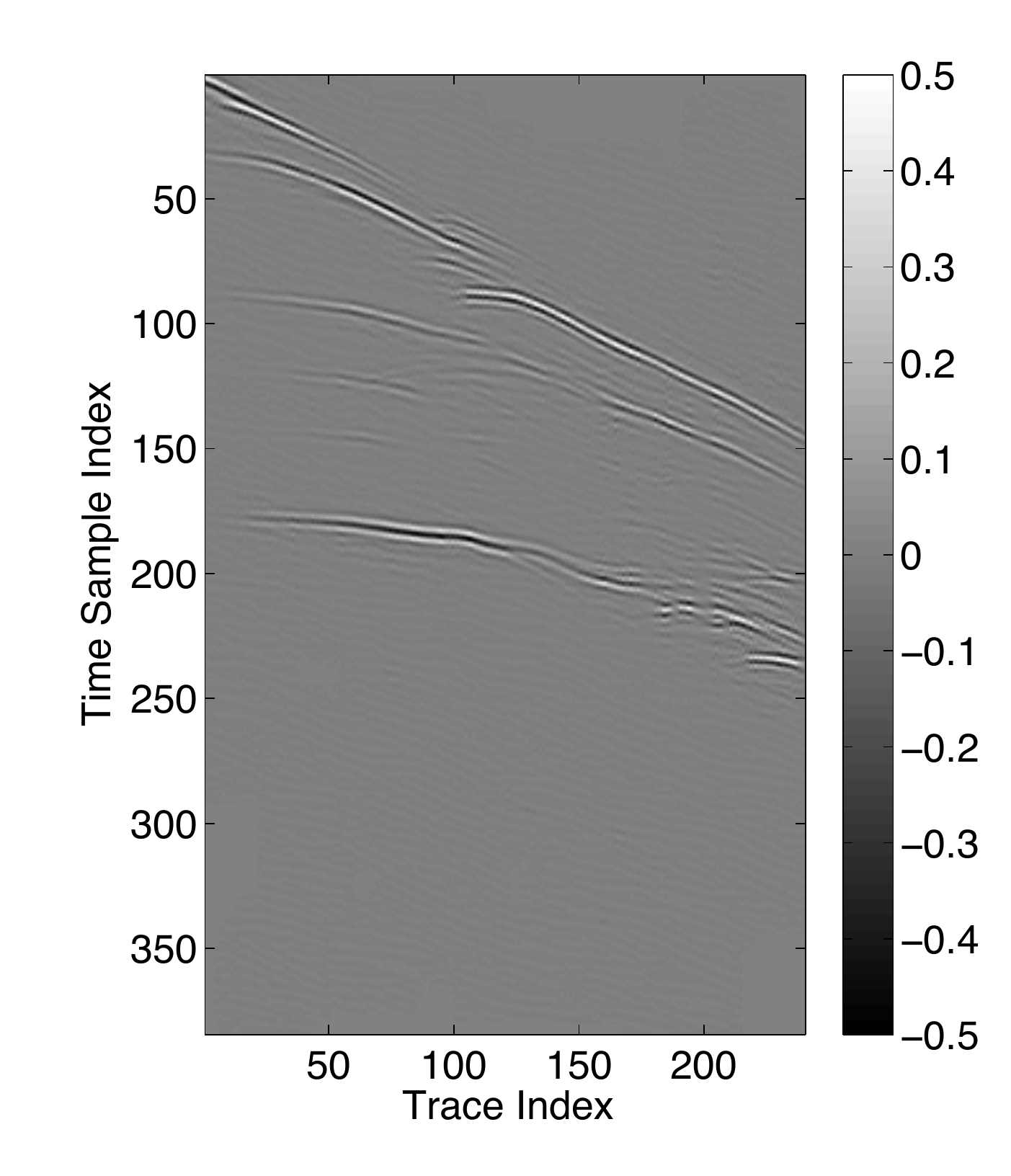}
		\end{minipage}
		\label{fig:bppublic_inpaintedData_sparseKsvd_dct_050}
	}
	\subfloat[Error Panel, 50\% missing]
	{
		\begin{minipage}{0.4\linewidth}
			\centering
			\includegraphics[width=\textwidth]{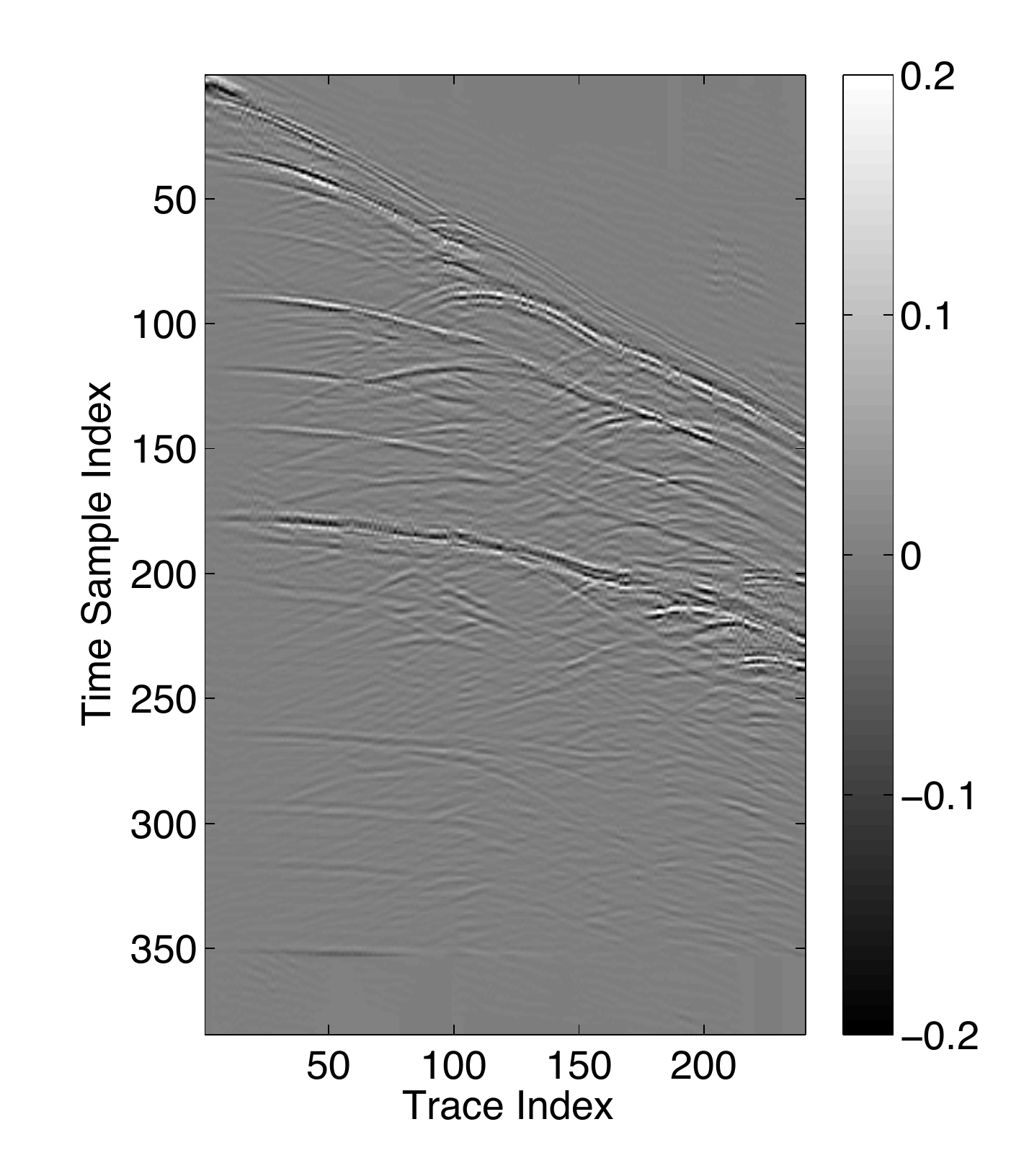}
		\end{minipage}
		\label{fig:bppublic_inpaintedDiffData_sparseKsvd_dct_050}
	}
	\caption{Recovery results of double-sparsity dictionary learning method using DCT matrix as the base dictionary from noisy datasets in Figure~\ref{fig:bppublic_data_for_inpainting}.
		(a) Recovery with 33\% missing traces, (b) difference between (a) and the original in Figure~\ref{fig:bppublic_data_for_inpainting}a, (c) recovery with 50\% missing traces, (d) difference between (c) and the original.   Note the change of gray scale for the error panels.}
	\label{fig:bppublic_inpainting_sparseKsvd_dct}
\end{figure}

Next, recovery experiments are carried out following the procedure in Algorithm \ref{alg:dataset-inpainting}. In the patch-based interpolation framework, one can fill ``holes'' whose sizes are smaller than that of the atoms \cite{Mairal:2008aa}. Therefore, to guard against clusters of missing traces, the patch size is set to a slightly larger size $n_z \times n_x = 24 \times 24$, and a non-redundant DCT dictionary $\mathbf{\Phi}$ of size $N \times N = 576 \times 576$ is selected as the base dictionary. The DCT basis elements are purely real and orthogonal so its computation is very efficient. Currently, the DCT is a widely adopted transform in many well known patch-based image processing schemes, e.g., JPEG and MPEG. Similarly, a total number of 10,000 overlapping patches are randomly selected from the corrupted seismic dataset for dictionary learning and the sparse matrix $\mathbf{A}$ is initialized to $\mathbf{I}_{576\times576}$. The atom sparsity level $p$ is set to 50. For the case with 33\% missing traces, Figure \ref{fig:dctAtoms_576x576} shows the non-redundant DCT base dictionary, while the learned sparse matrix $\mathbf{A}$ after $K = 20$ training iterations is visualized in Figure \ref{fig:bppublic_A_dct_forInpainting}. Based on our simulations, we notice that the essence of the dictionary learning is the ``learning" process. The choice of initial dictionary $\mathbf{\Phi}$ does not make a big difference. For example, when we initialize $\mathbf{\Phi}$ with a curvelet dictionary, the improvement in the reconstruction results is negligible. However, the computing time increases significantly, because of the redundancy of the curvelet dictionary. The overall dictionary, $\mathbf{D} = \mathbf{\Phi A}$ of size $576 \times 576$, is visualized in Figure \ref{fig:bppublic_D_dct_forInpainting}.

\begin{figure}[h!]
	\centering
	\subfloat[Fix $\sigma = 0.1$ and compare with traditional multiscale transforms]
	{
		\begin{minipage}{0.45\linewidth}
			\centering
			\includegraphics[width=\textwidth]{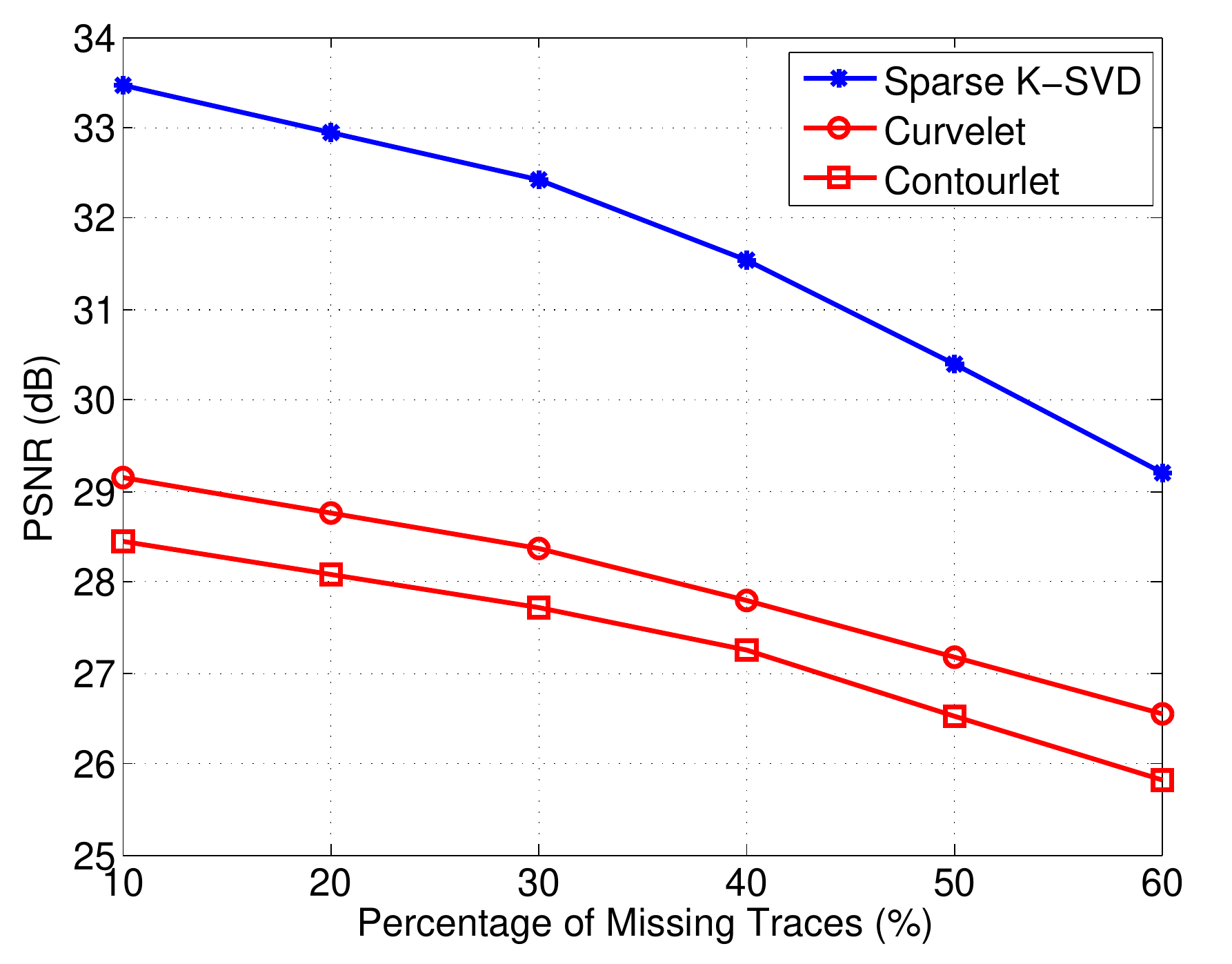}
		\end{minipage}
		\label{fig:bppublic_psnr_vs_nullratio}
	}
	\subfloat[Multiple $\sigma$ values]
	{
		\begin{minipage}{0.45\linewidth}
			\centering
			\includegraphics[width=\textwidth]{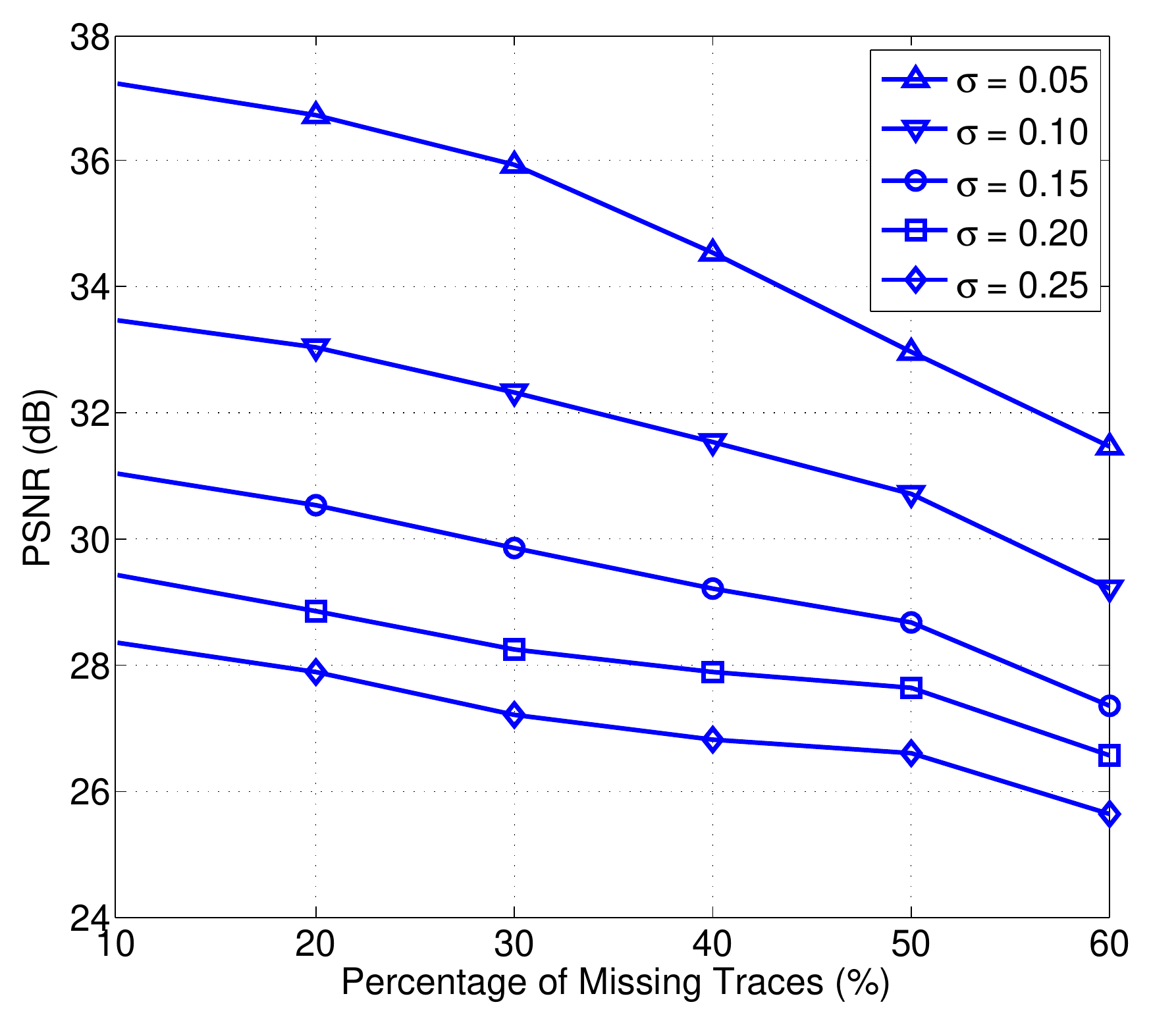}
		\end{minipage}
		\label{fig:bppublic_psnr_vs_nullratio_sigma}
	}
	\caption{PSNR versus percentage of missing traces for (a) different dictionaries and (b) different noise levels when using the double-sparsity K-SVD method.}
\end{figure}

Based on this double-sparsity learned dictionary, the recovery result can be obtained by (\ref{eq:dataset-inpainting-least-squares-solution}). Throughout this paper we define the PSNR as 
\begin{equation}
\mbox{PSNR} = 20\log \frac{\mathbf s_{\mbox{max}}\sqrt{N_x N_z}}{ \|\mathbf s -\hat{\mathbf{s}}\|},
\end{equation}
where $\mathbf s_{\mbox{max}}$ is the maximum possible value of the seismic data after normalization, and $N_x$, $N_z$ are the number of traces and time samples per trace, respectively.
The measured performance has been improved to PSNR = 32.11\,dB and PSNR = 30.31\,dB, as shown in Figure \ref{fig:bppublic_inpaintedData_sparseKsvd_dct_033} and \ref{fig:bppublic_inpaintedData_sparseKsvd_dct_050} for 33\% and 50\% missing traces, respectively. The corresponding error panels are shown in Figures \ref{fig:bppublic_inpaintedDiffData_sparseKsvd_dct_033} and \ref{fig:bppublic_inpaintedDiffData_sparseKsvd_dct_050}. When compared to the contourlet and curvelet transforms, the double-sparsity result exhibits no pseudo-Gibbs artifacts around the wave fronts. More experiments were performed in which the percentage of missing traces ranges from 10\% to 60\% and the PSNR performance curves are provided in Figure \ref{fig:bppublic_psnr_vs_nullratio}. The result with the double-sparsity dictionary learning method based on Algorithm \ref{alg:dataset-inpainting}, which is a modified version of the sparse K-SVD algorithm, yields significantly better PSNR values than the recovery with fixed transforms. In order to test the performance of the proposed scheme against noise, we also run the recovery simulation for different $\sigma$ values to obtain Figure \ref{fig:bppublic_psnr_vs_nullratio_sigma}. The double-sparsity recovery method provides correct and robust interpolation results for $\sigma$ up to 0.25, considering that the dynamic range of each trace is only 1 after normalization.

\section{Conclusions}\label{sec:conclusions}
For seismic datasets contaminated by random noise and missing traces, we presented a double-sparsity dictionary learning scheme to recover the data from these two types of distortions simultaneously. The main contribution of this work lies in the extension of the sparse K-SVD algorithm with a masking operator that tracks the missing data locations during the dictionary learning process. In addition, in order to solve the optimization involving the introduced masking operator, we adopt a weighted low-rank approximation algorithm to handle the dictionary updating. Numerical simulations on a benchmark dataset illustrate the validity of this new approach and its advantages over fixed transform approaches in the sense of yielding restoration with better PSNR and greatly reduced pseudo-Gibbs artifacts.

\end{document}